\documentclass[12pt]{iopart}
\usepackage{graphicx}
\usepackage{dcolumn}
\usepackage{bm}
\usepackage{color}
\usepackage{iopams}
\usepackage{multirow}
\usepackage{algorithm}
\usepackage{algorithmicx}
\usepackage{algpseudocode}
\usepackage{subfigure}
\usepackage{hyperref}
\usepackage{wasysym}
\usepackage{xcolor}
\usepackage{overpic}
\usepackage{color}
\usepackage{lineno}
\usepackage{caption}

\begin{document}

\title[]{A sequentially generated variational quantum circuit with polynomial complexity}

\author{Xiaokai Hou$^{1,2}$, Qingyu Li$^1$, Man-Hong Yung$^{2,3,4}$, Xusheng Xu$^2$, Zizhu Wang$^{1}$, Chu Guo$^1$ and Xiaoting Wang$^1$}

\address{$^1$ Institute of  Fundamental and Frontier Sciences, University of Electronic Science and Technology of China, Chengdu, Sichuan, 610051, China}
\address{$^2$ Central Research Institute, 2012 Labs, Huawei Technologies}
\address{$^3$ Department of Physics, Southern University of Science and Technology, Shenzhen, 518055, China}
\address{$^4$ Shenzhen Institute for Quantum Science and Engineering, Southern University of Science and Technology, Shenzhen, 518055, China}

\eads{\mailto{zizhu@uestc.edu.cn}, \mailto{guochu604b@gmail.com}, \mailto{xiaoting@uestc.edu.cn}}
\vspace{10pt}

\begin{abstract}
Variational quantum algorithms have been a promising candidate to utilize near-term quantum devices to solve real-world problems.
The powerfulness of variational quantum algorithms is ultimately determined by the expressiveness of the underlying quantum circuit ansatz for a given problem.
In this work, we propose a sequentially generated circuit ansatz, which naturally adapts to $1$D, $2$D, $3$D quantum many-body problems. 
Specifically, in $1$D our ansatz can efficiently generate any matrix product states with a fixed bond dimension, while in $2$D our ansatz generates the string-bond states.
As applications, we demonstrate that our ansatz can be used to accurately reconstruct unknown pure and mixed quantum states which can be represented as matrix product states, and that our ansatz is more efficient compared to several alternatives in finding the ground states of some prototypical quantum many-body systems as well as quantum chemistry systems, in terms of the number of quantum gate operations.
\end{abstract}

%
\maketitle
%
%

\section{Introduction}
Fueled by the advances of quantum technologies, quantum computing has grown rapidly and entered a stage which is the so-called noisy intermediate-scale quantum devices~(NISQ) era~\cite{Preskill2018quantum}. Although there is still a long way to achieve fully fault-tolerant quantum computing, various quantum algorithms have already been demonstrated on near term quantum devices, ranging from random quantum circuit sampling~\cite{arute2019quantum,PhysRevLett.127.180501,ZhuPan2021}, Boson sampling~\cite{zhong2020quantum,madsen2022quantum}, prime factorization~\cite{365700} to quantum walk~\cite{childs2003exponential}, solving linear equations~\cite{PhysRevLett.103.150502,doi:10.1137/16M1087072} and machine learning~\cite{PhysRevLett.109.050505,PhysRevLett.113.130503,lloyd2014quantum,doi:10.1126/science.abn7293,abbas2021power}. Specially, as one of the most promising quantum algorithms to achieve practical quantum advantage, variational quantum algorithms~(VQAs) have attracted tremendous attentions for their broad applications in computational chemistry~\cite{PhysRevX.6.031007,kandala2017hardware,mccaskey2019quantum,mcardle2020quantum,HuangLong2022}, dynamical quantum simulation~\cite{PhysRevX.7.021050,PhysRevLett.125.010501,mcardle2019variational,PRXQuantum.2.030307,zhang2020low}, quantum error correction~\cite{johnson2017qvector,PhysRevApplied.15.034068,cao2022quantum}, quantum generative models~\cite{PhysRevA.98.012324,benedetti2019generative,PhysRevResearch.2.033125} and quantum neural networks~\cite{PhysRevA.101.032308,PhysRevA.98.032309,hou2021universal,cong2019quantum}. Among them, the variational quantum eigensolver~(VQE) has been proposed for efficiently approximating the ground energy of a given Hamiltonian~\cite{kandala2017hardware,peruzzo2014variational,PhysRevLett.122.230401,aspuru2005simulated} and has been realized on several quantum devices, such as ion traps~\cite{meth2022probing,PhysRevX.8.031022,PhysRevA.95.020501}, photonic chips~\cite{peruzzo2014variational,Lee:22}, nuclear magnetic resonance systems~\cite{li2011solving}, and superconducting quantum devices~\cite{PhysRevX.6.031007,kandala2017hardware,PhysRevX.8.011021}. 

As one typical hybrid quantum-classical algorithm, a  VQE comprises a quantum simulator and a classical optimizer. Through iteratively updating parameters in the simulator with the classical optimizer, we can minimize the average energy of a given Hamiltonian $H$ by using the gradient descent algorithm~\cite{Sweke2020stochasticgradient,suzuki2021normalized}, and finally obtain the ground energy. Despite wide applications of VQE, two fundamental problems still remain open and make it challenging to understand the effectiveness of VQE. One question is the ground state of which kinds of quantum systems can be efficiently approximated by a VQE with polynomial circuit complexity. The other question is how to avoid the barren plateau~\cite{mcclean2018barren,wang2021noise,Arrasmith2021effectofbarren} during the optimization of VQE such that it can converge to the desired ground energy. For the latter question, several established methods have demonstrated that the effect of the barren plateau can be weakened by using the classical shadows~\cite{PRXQuantum.3.020365}, the adaptive, Problem-Tailored (ADAPT)-VQE ansatz~\cite{grimsley2022adapt}, an random initialization strategy~\cite{Grant2019initialization}, and so on. For the former question, it relies on the structure of a variational quantum circuit used in a VQE. The structure of a variational quantum circuit, known as the circuit ansatz, determines the quantum states generated by a VQE. A VQE without a carefully designed ansatz will fail to approximate the ground state of a given Hamiltonian. Although there have been several ansatz proposed for the molecule Hamiltonian and the unconstrained-optimization-problem Hamiltonian, i.e. the hardware-efficient~(HE) ansatz~\cite{kandala2017hardware}, Unitary Coupled clustered ansatz~\cite{doi:10.1021/acs.jctc.8b01004}, quantum alternating operator ansatz~\cite{a12020034} and so on, there still exists a wide range of quantum systems need to be solved.

In this work, we address this challenge by proposing a sequentially generated~(SG) parameterized quantum circuit ansatz, which easily adapts to the $1$D, $2$D and $3$D quantum many-body systems. In the $1$D case we show that our ansatz can generate any matrix product state (MPS) with a fixed bond dimension using a polynomial number of gates, and we demonstrate two applications which show the accuracy and efficiency (in terms of the number of gates) of our ansatz including 1) reconstructing unknown pure or mixed quantum states which are assumed to be able to be represented as MPSs and 2) searching for ground states of $1$D quantum systems and quantum chemistry systems using VQE based on our ansatz. In the $2$D case our ansatz can generate certain string-bond states~\cite{SchuchCirac2008,GlasserCirac2018} and we demonstrate that our ansatz could be used to accurately approximate the ground state the $2$D Ising model with size up to $5\times 5$ with a low depth. The effectiveness of our ansatz for computing the ground state of the $3$D quantum Ising model is also considered in the end. Our results demonstrate that one could design parameterized quantum circuit ansatz that are inspired from the well studied tensor network state ansatz for variational quantum algorithms, such that one can largely benefit from the success of the later.

\section{Variational quantum eigensolver and sequentially generated ansatz}\label{Sec:SG_ansatz}

In this section, we briefly review VQE and then show the structure of SG ansatz in detail. 

\begin{figure}[tp]
  \centering
  \includegraphics[width=0.6\textwidth]{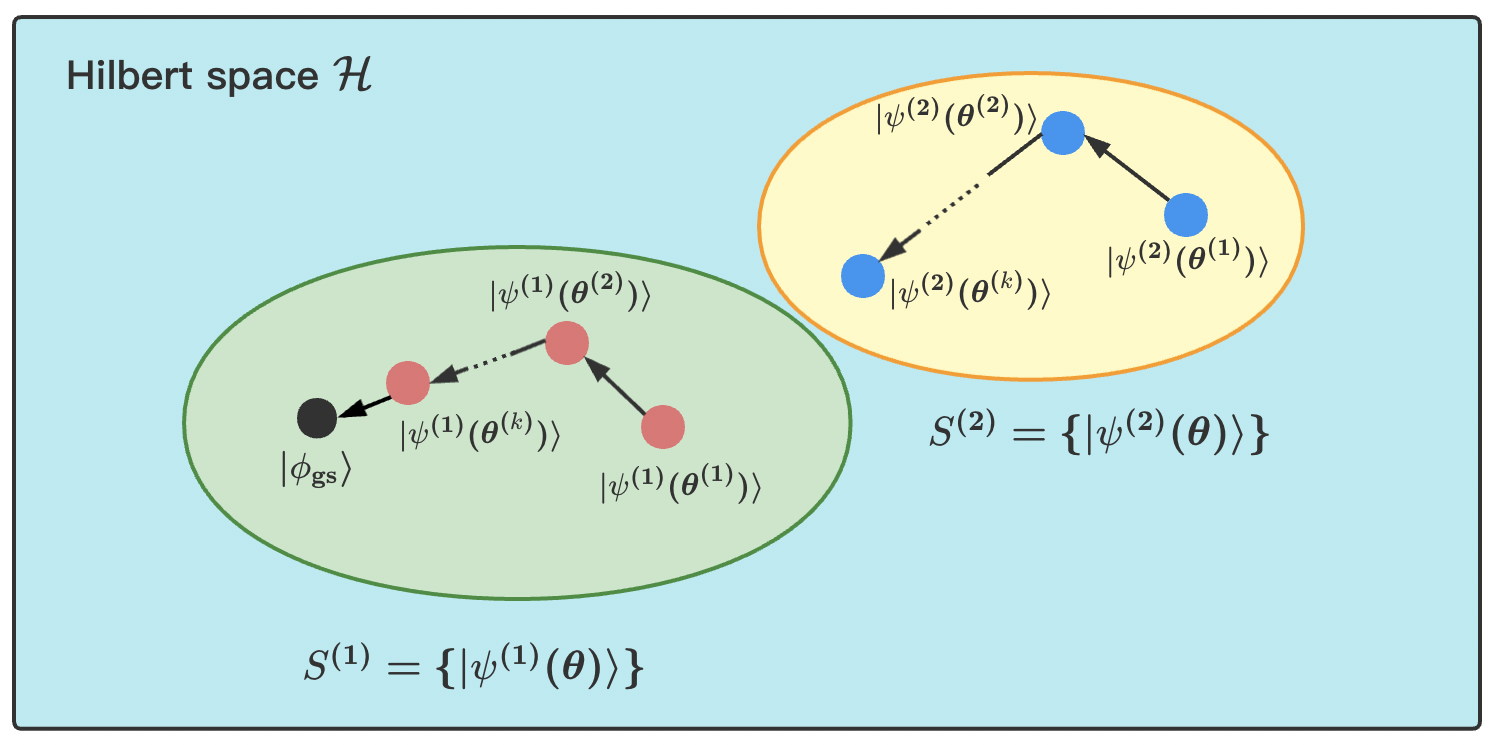}
  \caption{An illustration for the optimization of a VQE. For different circuit ansatz $U^{(j)}$, it determines a set of variational quantum state $S^{(j)}=\{|\psi^{(j)}(\bm{\theta})\rangle\}$, which is shown as the colored circle. The black point indicates the ground state of a given Hamiltonian. The red points and the blue points represent the variational quantum states based on different $U^{(j)}(\bm{\theta})$. With an initialization of parameters $\bm{\theta^{(1)}}$, the optimization of each ansatz is shown as the directed line.
   }
  \label{set_variatinal_state}
\end{figure}

\subsection{Variational quantum eigensolver}
As one of the typical variational quantum algorithms, VQE utilizes a quantum simulator and a classical optimizer to approximate the ground state $|\phi_{\rm{gs}}\rangle$ of a given Hamiltonian $H$ through an iterative optimization process on a hybrid quantum-classical computer. At the $j$-th iteration, we apply a parameterized quantum circuit $U(\bm{\theta}^{(j)})$ to an initial state $|\psi_0\rangle$, followed by measuring the average energy of the quantum simulator as 
\begin{equation}\label{E_theta}
    \langle H(\bm{\theta}^{(j)})\rangle\equiv\langle\psi_0|U^{\dagger}(\bm{\theta}^{(j)})HU(\bm{\theta}^{(j)})|\psi_0\rangle
\end{equation}
where $\bm{\theta}$ indicate the angles of local rotation gates in a variational quantum circuit. With the classical optimizer, we can update the parameters $\bm{\theta}^{(j)}\rightarrow \bm{\theta}^{(j+1)}$ by minimizing equation~(\ref{E_theta}) through a gradient-based optimization method, such as the gradient descent method~\cite{Sweke2020stochasticgradient,9605301}, the quasi-Newton method~\cite{liu1989limited} and the Adam method~\cite{kingma2014adam}. After multiple iterations, we can obtain the optimal $\bm{\theta^*}$ such that the average energy $\langle H(\bm{\theta^*})\rangle$ is close to the ground energy of $H$ and the variational quantum state $|\psi(\bm{\theta^*})\rangle\equiv U(\bm{\theta^*})|\psi_0\rangle$ can well approximate $|\phi_{\rm{gs}}\rangle$. 

In general, a variational circuit ansatz $U(\bm{\theta})$ determines a set of variational quantum states $S=\{|\psi(\bm{\theta})\rangle\}$. The optimization of VQE can be considered as a process to find an optimal variational state $|\psi(\bm{\theta}^*)\rangle\in S$, and the state is a good approximation to the ground state $|\phi_{\rm{gs}}\rangle$ of $H$. As illustrated in figure~(\ref{set_variatinal_state}), for the effectiveness of VQE, it is necessary to design a variational circuit ansatz $U(\bm{\theta})$ satisfying $|\phi_{\rm{gs}}\rangle\in S$. A VQE will fail if the optimal state in its variational quantum state set can not approximate the ground state efficiently.

\subsection{Sequentially generated circuit ansatz}
As an alternative variational quantum circuit ansatz, the SG ansatz consists of multiple variational quantum circuit blocks, each of which is a parametrized quantum circuit applied to several adjacent qubits. With such a structure, the SG ansatz naturally adapts to quantum many-body problems. Specifically, for $1$D quantum systems, the SG ansatz can efficiently generate any matrix product states with a fixed bond dimension. For $2$D systems, the SG ansatz can generate string-bond states. The details of SG ansatz in each case of quantum systems are introduced as follows.

\subsubsection{SG ansatz for 1D system}

For 1D quantum systems, the SG ansatz aims to generate a variational quantum state $|\psi(\bm{\theta})\rangle$, which can be efficiently characterized by an MPS. The formula of a $n$-qubit MPS, $|\Psi\rangle \in \mathbb{C}_2^{\otimes n}$, is given as 
\begin{equation}
    |\Psi\rangle=\sum_{s_1,s_2,\cdots,s_n}\Tr[A_1^{(s_1)}A_2^{(s_2)}\cdots A_n^{(s_n)}]|s_1s_2\cdots s_n\rangle
\end{equation}
where $s_j\in\{0,1\}$ and $A_j^{(s_j)}$ is a complex matrix. We define $R\equiv \max({\rm rank}(A_j^{(s_n)}))$ as the \textit{bond dimension} of $|\Psi\rangle$ which is determined by the quantum entanglement of $|\Psi\rangle$. $R$ is larger if the quantum state is more entangled. For a $1$D gapped local Hamiltonian, the area law guarantees that the entanglement of its ground state is approximately constant~\cite{hastings2007area}. 

\begin{figure}[tp]
    \begin{center}
        \includegraphics[width=0.9\linewidth]{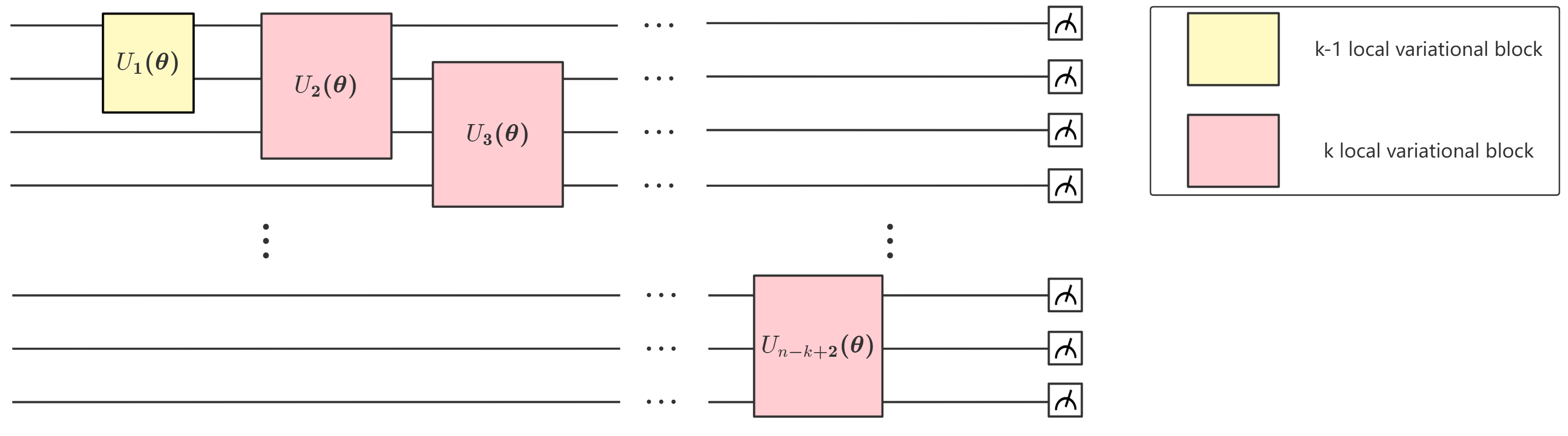}
    
        \caption{The SG ansatz to generate a $n$-qubit MPS state with bond dimension $R=4$. The hyperparameter $k$ is equal to $3$. The yellow rectangle indicates a $(k-1)$-local variational circuit block consisting of several single-qubit and two-qubit gates. The red rectangle indicates a $k$-local variational circuit block.}
        \label{Fig1}
    \end{center}
\end{figure}

\begin{figure}[htp]
    \centering
    \includegraphics[width=0.4\linewidth]{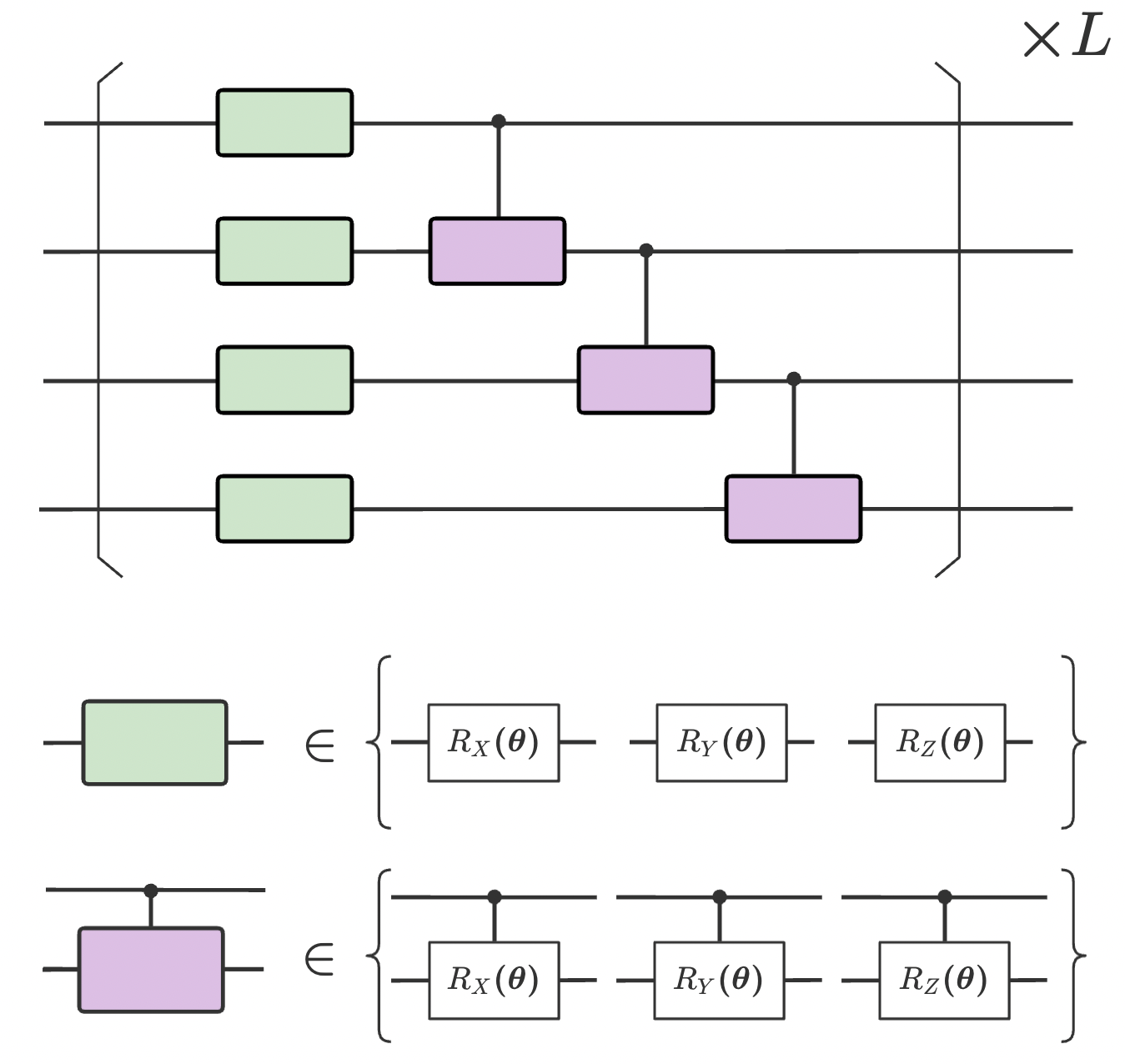}
    \caption{A $4$-qubit variational circuit block of the SG ansatz consisting of $L$ layers. In each layer, the single qubit rotation gate is chosen from the set $\{R_X(\theta), R_Y(\theta), R_Z(\theta)\}$ and the two-qubit rotation gate is chosen from $\{CR_X(\theta), CR_Y(\theta), CR_Z(\theta)\}$
    }
    \label{SG_layer}
\end{figure}

To generate a $n$-qubit MPS state with bond dimension $R$, we firstly calculate a hyperparameter $k=\lceil \log(R) \rceil+1$, then place a $(k-1)$-local variational circuit block and $n-k+1$ $k$-local variational circuit blocks into the SG ansatz, as shown in figure~(\ref{Fig1}). Inspired by Ref.~\cite{cramer2010efficient}, the block-place rule is chosen as: i) number the qubits from top to bottom as $1$ to $n$, ii) place a $(k-1)$-local variational circuit block on the qubit ranging from $1$ to $k-1$, iii) place $n-k+1$ k-local variational circuit blocks, where the $j$th block acts on the qubits ranging from $j$ to $j+k-1$ with $j=1,2,\cdots,n-k+1$. Specially, each local variational block consists of $L$-layer circuits as shown in figure~(\ref{SG_layer}). In each layer, we firstly apply single qubit rotation gates to each qubit and then apply several two-qubit gates. For the single qubit rotation gate, we randomly choose one from a set $\{R_X(\theta), R_Y(\theta), R_Z(\theta)\}$ where $X$, $Y$ and $Z$ indicate the rotation through angle $\theta$ around the $x$-axis, $y$-axis and $z$-axis, and for the two-qubit gate, we randomly choose one from a set $\{CR_X(\theta), CR_Y(\theta), CR_Z(\theta)\}$. 

Based on above structure, we can analyze the circuit complexity of the SG ansatz to generate a $n$-qubit MPS, $|\Psi\rangle$, with bond dimension $R$. According to Ref.~\cite{cramer2010efficient}, for a given $|\Psi\rangle$, one can disentangle it into $|0\rangle^{\otimes n-k+1}\otimes|\eta\rangle$ by sequentially applying $O(n)$  $k$-local unitary matrices, and further applies one $k-1$ local unitary matrix to transform $|\eta\rangle$ into $|0\rangle^{\otimes k-1}$ where $k=\lceil\log_2R\rceil+1$. To construct such $|\Psi\rangle$ from $|0\rangle^{\otimes n}$, we can train one $k-1$ local variational block and $O(n)$ $k$-local variational blocks such that these blocks can achieve the inverse process of disentanglement. Theoretically, we require $O(4^k)$ quantum gates for each block to approximate an arbitrary unitary matrix. Hence the circuit complexity for the SQ ansatz to generate the desired MPS state is $O(nR^2)$.

\subsubsection{SG ansatz for 2D and 3D system}
\begin{figure}[htb]
    \centering
    \subfigure[]
    {
        \begin{minipage}{.23\linewidth}\label{line1}
        \centering
        \includegraphics[width=1\textwidth]{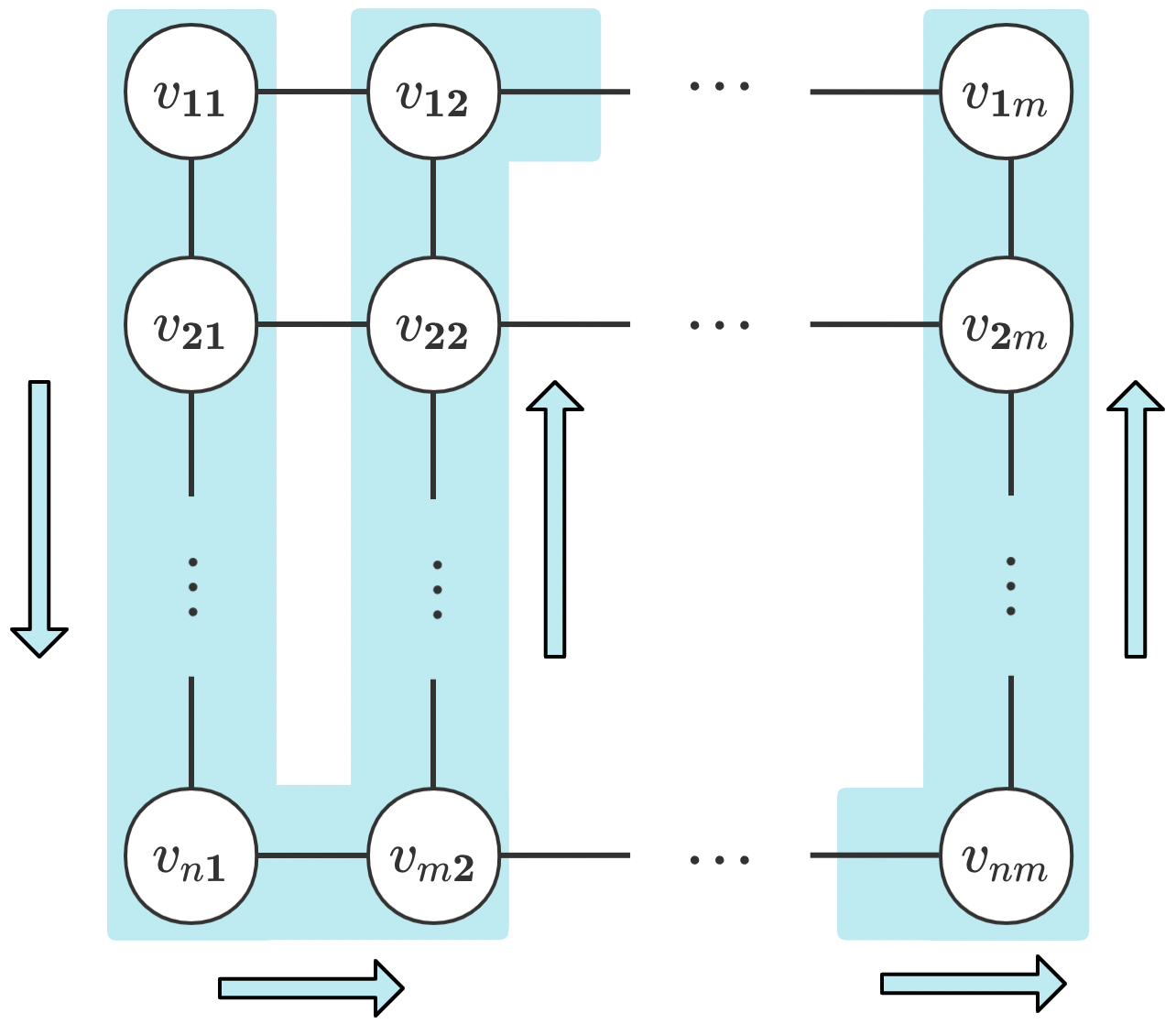}
        \end{minipage}
    }
    \subfigure[]
    {
        \begin{minipage}{.23\linewidth}\label{line2}
        \centering
        \includegraphics[width=0.9\textwidth]{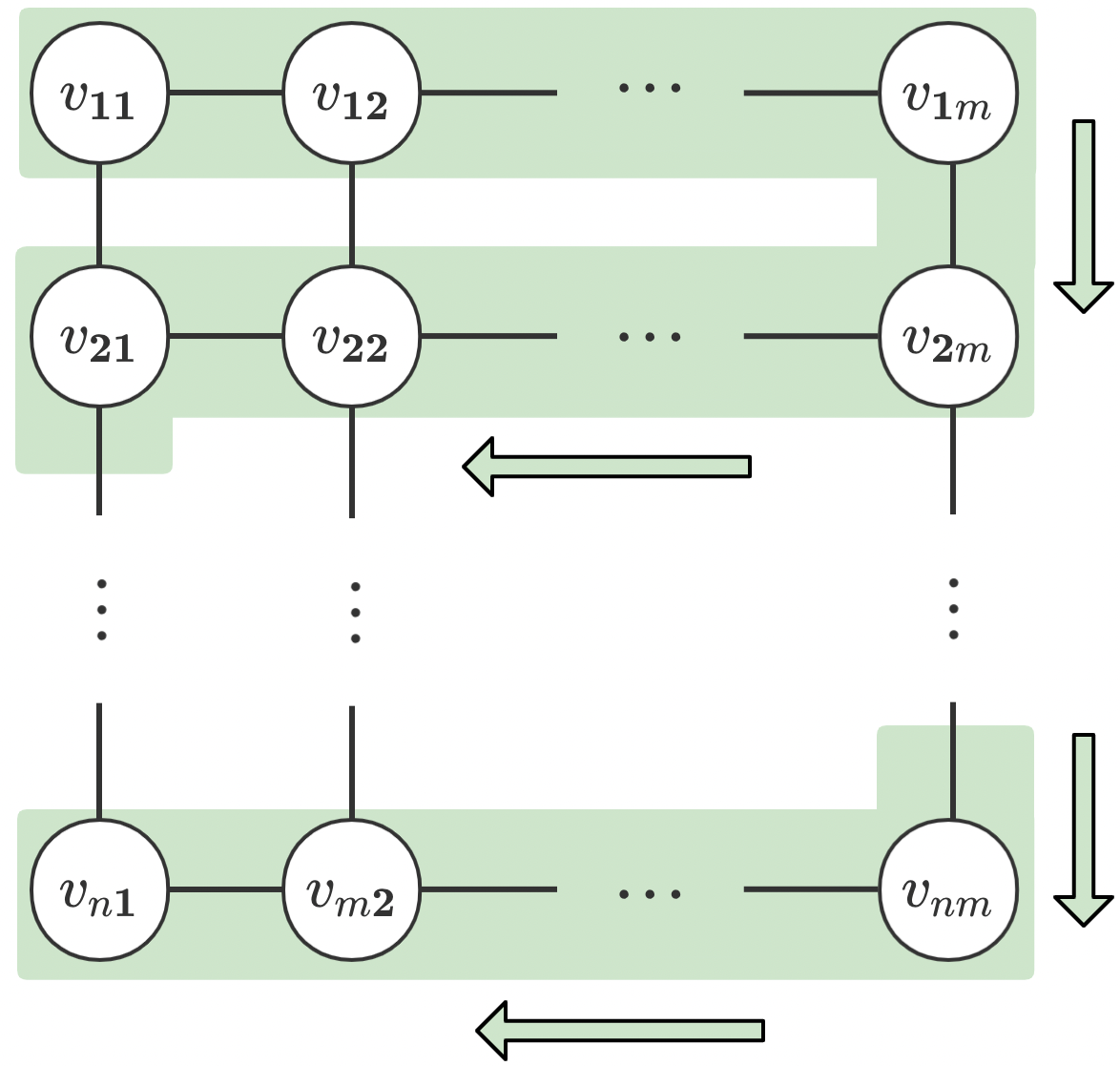}
        \end{minipage}
    } 
    \\
    \subfigure[]
    {
        \begin{minipage}{.45\linewidth}\label{2D_circuit}
        \centering
        \includegraphics[width=1\textwidth]{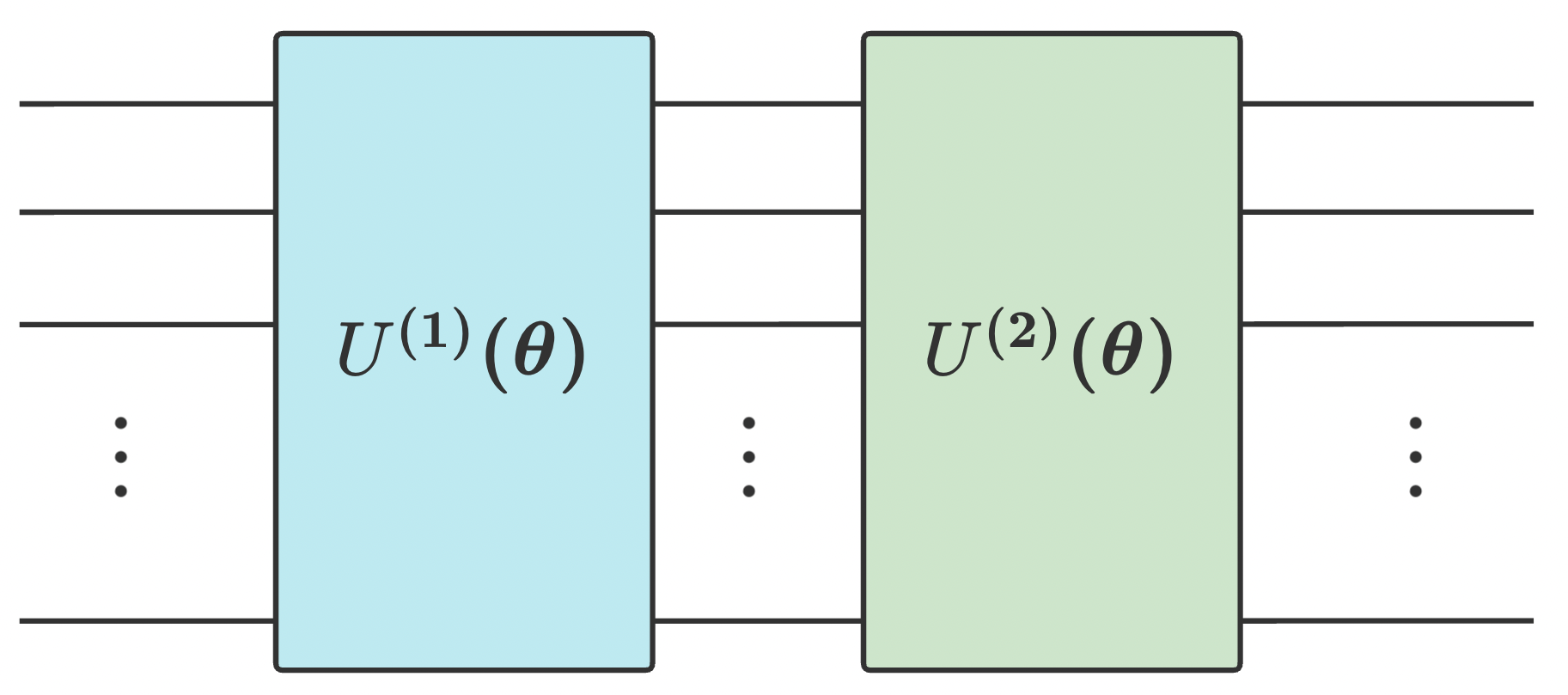}
        \end{minipage}
    }

    \caption{SG ansatz for solving 2D quantum model.  \textbf{(a)} The column-orientated line to organize qubits. \textbf{(b)} The row-orientated line to organize qubits. \textbf{(c)} 2D SG ansatz. $U^{(1)}(\bm{\theta})$ and $U^{(2)}(\bm{\theta})$ are the circuits generated by using the column-orientated line and the row-orientated line.
    }
\end{figure}

For the 2D quantum model, considering a $N$-qubit lattice which has $n$ rows and $m$ columns, we utilize two lines of qubits to construct the SG ansatz. As shown in figure~\ref{line1}, the first line is column-orientated. This line starts from the qubit in the upper left corner, follows the column to the qubit in the lower left corner, and then begins from the qubit at the bottom of the second column and goes up. It ends after traversing all the qubits in the lattice. The second line, as shown in figure~\ref{line2}, is row-orientated. It also starts at the top left qubit but follows the row to the top right qubit, then begins at the rightmost qubit in the second row and goes left. It ends after traversing all the qubits in the lattice. After obtaining these two lines, we choose a suitable bond dimension $R$ and construct two circuits as introduced in the $1$D case, where the order of qubits is organized using the two lines. Marking the circuit generated by using the column-orientated line as $U^{(1)}(\bm{\theta})$ and the circuit generated by using the row-orientated line as $U^{(2)}(\bm{\theta})$, the SG ansatz used for solving 2D quantum model is shown in figure~\ref{2D_circuit}. By construction, our ansatz generates a specific type of string-bond state in which each MPS extends to the whole system~\cite{SchuchCirac2008}.

\begin{figure}[tp]
\centering

\includegraphics[width=0.7\linewidth]{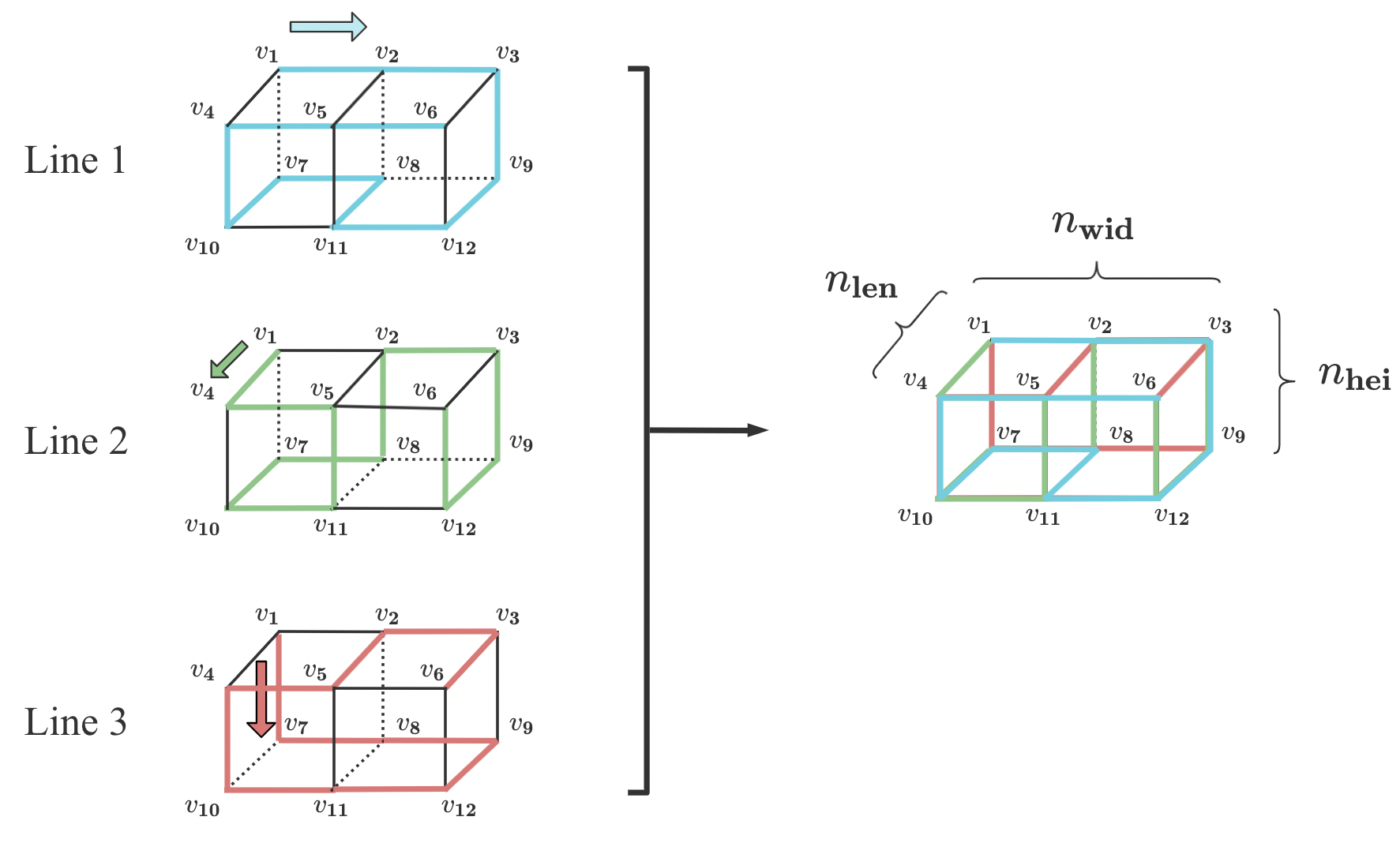}

\caption{An illustration for generating line sets for a $3$D quantum Ising model with $n_{len}=2$, $n_{wid}=3$ and $n_{hei}=2$. Each vertex $v_{ij}$ represents one qubit and each edge represents the nearest-neighbor interaction. Each line starts from vertex $v_1$ and transverses all of the vertices in the cube, finally ending at vertex $v_6$. Combining all three lines, we can find all of the edges in the cube are traversed. }
\label{3D_model}
\end{figure}

For the 3D quantum model, we mainly focus on the 3D Ising cube model as shown in figure~\ref{3D_model} where each vertex $v_i \in V=\{v_j\}$ indicates one qubit and each edge $e_{ij} \in E=\{e_{ij}=(v_i, v_j)\}$ represents the nearest neighbor interaction between $v_i$ and $v_j$. $V$ and $E$ respectively represent the set of all vertexes and edges in a cube. Similar to the circuit for the 2D case, the first step to construct the SG ansatz for a 3D model is to find a suitable set of lines such that each edge can be contained in the line set. We generate the line set by using the following method: i) choose $v_0$ as the start of all lines, ii) for all $v_j \in V/v_0$, take $v_j$ as the destination and use the depth-first search method to find the set of all paths, $P^{(j)}=\{p_k^{(j)}\}$, each of which should traversal all of the vertexes, iii) choose the vertex $v_j=\max_{j} |P^{(j)}|$ as the destination of the lines iv) choose the $L$ paths such that these paths contain all of the edge and take them as the set of lines. Next, We choose a suitable bond dimension $R$ and construct a circuit $U^{(i)}(\bm{\theta})$ for each line $p_i^{(j)}\in \{p_1^{(j)},p_2^{(j)},\cdots,p_L^{(j)}\}$. Combining these circuits, we finally construct the circuit $U=\Pi_{i=1}^L U^{(i)}(\bm{\theta})$ for the 3D quantum Ising cube.

\section{SG ansatz for reconstructing unknown quantum states}\label{sec:qst}
As mentioned in the previous section, the SG ansatz with polynomial circuit complexity has the ability to generate a MPS with a fixed bond dimension. 
To demonstrate the powerfulness of our ansatz, we first apply it for reconstruct unknown pure and mixed quantum states using a variational quantum algorithm as in Ref.~\cite{LiuWu2020}, in which the quantum fidelity between the reconstructed quantum state generated by applying an SG circuit $U(\vec{\theta})$ to an initial state $|0\rangle$ and the unknown quantum state. In case the unknown state is a pure state written as $\vert \psi\rangle$, the loss function is
\begin{equation}
    L(\bm{\theta}) = 1-F(|\psi\rangle,|\phi(\bm{\theta})\rangle)\equiv 1-|\langle\psi|\phi(\bm{\theta})\rangle|^2,
\end{equation}
where $|\phi(\bm{\theta})\rangle=U(\bm{\theta})|0\rangle$.

\begin{figure}[tp]
\centering
\setlength{\belowcaptionskip}{-.3cm}
    \subfigure[]
    {
     \begin{minipage}{.4\linewidth}
     \centering
      \includegraphics[width=1\textwidth]{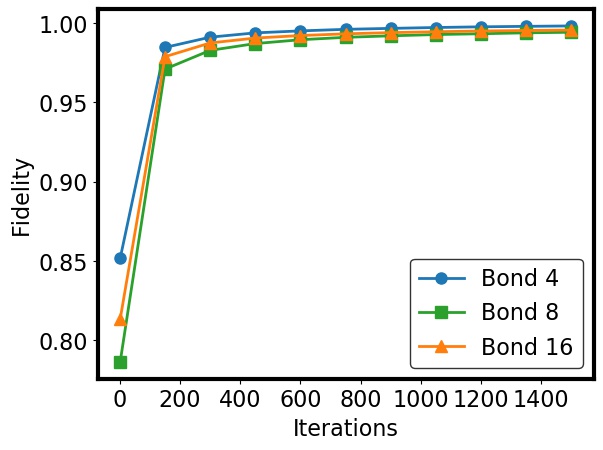}
      \label{Fig:pure_state_generation_n8}
     \end{minipage}
    }
    \subfigure[]
    {
     \begin{minipage}{.4\linewidth}
     \centering
      \includegraphics[width=1\textwidth]{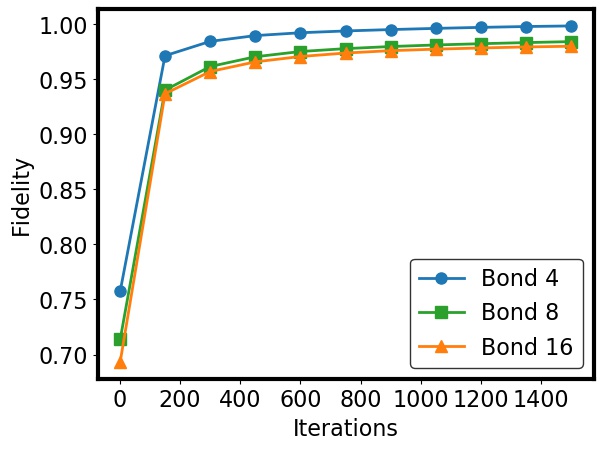}
      \label{Fig:pure_state_generation_n10}
     \end{minipage}
    } 
   \caption{Reconstructing unknown pure quantum states using a variational quantum algorithm based on our SG ansatz, in which each block is a $10$-layer circuit.  \textbf{(a)} Fidelities for reconstructing $8$-qubit quantum states which can be written MPSs with the bond dimensions $4$, $8$, $16$ respectively. \textbf{(b)} Fidelities for reconstructing $10$-qubit quantum states which can be written MPSs with bond dimensions $4$, $8$, $16$ respectively.}
   \label{Fig:pure_state_generation}
\end{figure}

In our simulation, we generate a synthetic dataset which are a set of $n$-qubit pure quantum states $\{\rho^{(i)}=|\psi^{(i)}\rangle\langle\psi^{(i)}|\}_{i=1}^{20}$, we have also considered different bond dimensions $R=4$, $8$ and $16$ for a $8$-qubit quantum state and a $10$-qubit quantum state, respectively. We set the number of layers in the SG ansatz to be $10$ and use the Adam optimization method to train the parameters. Taking the average fidelity of $20$ random target states as a function of the optimization iterations, we plot the results in figure~\ref{Fig:pure_state_generation}. figure~\ref{Fig:pure_state_generation_n8} is obtained based on $8$-qubit pure states with bond dimension $R=4$, $8$, $16$. After the optimization process, the average fidelity of the target states with $R=4$, $8$, $16$ is $99.83\%$, $99.43\%$, $99.57\%$. figure~\ref{Fig:pure_state_generation_n10} is obtained based on $10$-qubit pure states with bond dimension $R=4$, $8$, $16$. After the optimization process, the average fidelity of the target states with $R=4$, $8$, $16$ is $99.80\%$, $98.38\%$, $97.96\%$. From these results, we can see that our SG ansatz can be used to accurately reconstruct unknown pure quantum states that can be written as MPSs.  

Besides the pure states, we further use our SG ansatz to reconstruct mixed quantum states. 
To effectively implement the reconstruction process, we need to generate a purified quantum state using the SG ansatz for a given mixed state $\rho$. Assuming that the mixed state $\rho$ satisfies
\begin{equation}\label{eq:rho}
    \rho=\Tr_{n+1,\cdots,n+\log_2 M}(|\psi_A\rangle\langle\psi_A|).
\end{equation}
where each $|\psi_A\rangle\in\mathbb{C}_2^{\otimes n+\log_2 M}$ is a pure state. For the purification of $\rho$, we use a $(n+\log M)$-qubit SG ansatz $U(\bm{\theta})$ which is designed based on the bond dimension $k=\lceil \log_2 MR\rceil+1$. After applying $U(\bm{\theta})$ to an initial state $|0\rangle$ by using $|\phi(\bm{\theta})_A\rangle=U(\bm{\theta})|0\rangle$, we obtain the output according to
\begin{equation}
    \sigma_{\bm{\theta}}=\Tr_{n+1,\cdots,n+\log_2 M}(|\phi(\bm{\theta})_A\rangle\langle\phi(\bm{\theta})_A|),
\end{equation}
and measure the fidelity between $\sigma_{\theta}$ and the target mixed state $\rho$ by using the definition
\begin{equation}
    F(\rho,\sigma_{\bm{\theta}})\equiv \frac{|\Tr(\rho\sigma_{\bm{\theta}})|}{\sqrt{\Tr(\rho^2)}\sqrt{\Tr(\sigma_{\bm{\theta}}^2)}}.
\end{equation}
Similar to reconstructing pure quantum states, we take $L(\bm{\theta})=1-F(\rho,\sigma_{\bm{\theta}})$ as the loss function and minimize it to obtain the optimal parameters $\bm{\theta}^*$ through optimization.

\begin{figure}[tp]
\centering
\setlength{\belowcaptionskip}{-.5cm}
    \subfigure[]
    {
     \begin{minipage}{.4\linewidth}\label{mix_N4}
     \centering
      \includegraphics[width=1\textwidth]{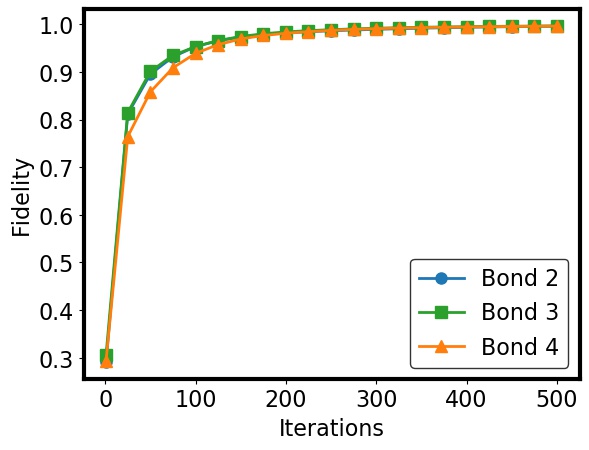}
     \end{minipage}
    }
    \subfigure[]
    {
     \begin{minipage}{.4\linewidth}\label{MIX_N6}
     \centering
      \includegraphics[width=1\textwidth]{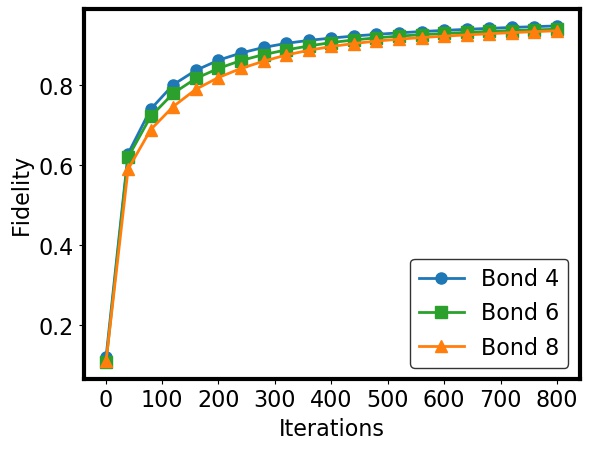}
     \end{minipage}
    } 

   \caption{Reconstructing unknown mixed quantum states with our SG ansatz, in which each block is a $10$-layer variational circuit.  \textbf{(a)} Reconstructing $4$-qubit mixed states for which $M=10$ and each pure state in equation(\ref{eq:rho}) can be written as MPSs with bond dimension $2$, $3$, $4$ respectively. \textbf{(b)} Reconstructing $6$-qubit mixed states for which $M=10$ and each pure state can be written as MPSs with bond dimension $4$, $6$, $8$ respectively.}
   \label{mix_state_generation}
\end{figure}

In our numerical simulation we randomly generate a set of $n$-qubit mixed states, $\{\rho^{(i)}=\sum_{j=1}^Mp^{(i)}_j|\psi^{(i)}_j\rangle\langle\psi^{(i)}_j|\}_{i=1}^{20}$ as the target states where $|\bm{p}^{(i)}|_1=1$ and $|\psi^{(i)}_j\rangle \in\mathbb{C}_2^{\otimes n}$. We mention that each $|\psi^{(i)}_j\rangle$ is a pure quantum state which can be depicted by an MPS with the bond dimension not greater than $R$. Furthermore, we assume the Schmidt number, $M$, is polynomial to the number of qubits. Hence, the bond dimension of each $\rho^{(i)}$ is less than $MR$, which allows us to design an efficient SG ansatz.
We implement the simulations based on $4$-qubit and $6$-qubit mixed states. For the $4$-qubit mixed states, we set the Schmidt number $M=10$ and adopt different bond dimensions $R=2$, $3$, $4$ to generate the target mixed states. Taking the average fidelity of $20$ random target mixed states as a function of the optimization iterations, we summarize the results in figure~\ref{mix_N4}. We can see that, after several iterations, the average fidelity of the target mixed states with $R=2$, $3$, $4$ is $99.62\%$, $99.68\%$, $99.65\%$. Furthermore, for $6$-qubit mixed states, we set the Schmidt number $M=10$ and adopt bond dimensions $R=4$, $6$, $8$ to generate the target mixed states. The results for $6$-qubit mixed states are shown in figure~\ref{MIX_N6}. After the optimization, the average fidelity of the target mixed states with $R=4$, $6$, $8$ is $94.89\%$, $93.91\%$, $93.58\%$. From these results, we can see that 
our SG ansatz is effective for reconstructing mixed quantum states which can be written as MPSs.

\section{SG ansatz for variational quantum eigensolver}\label{sec:vqe}
\begin{figure}[htp]
    \setlength{\belowcaptionskip}{-.3cm}
    \subfigure[]
    {
        \begin{minipage}{0.3\linewidth}
        \centering
        \includegraphics[width=1\textwidth]
        {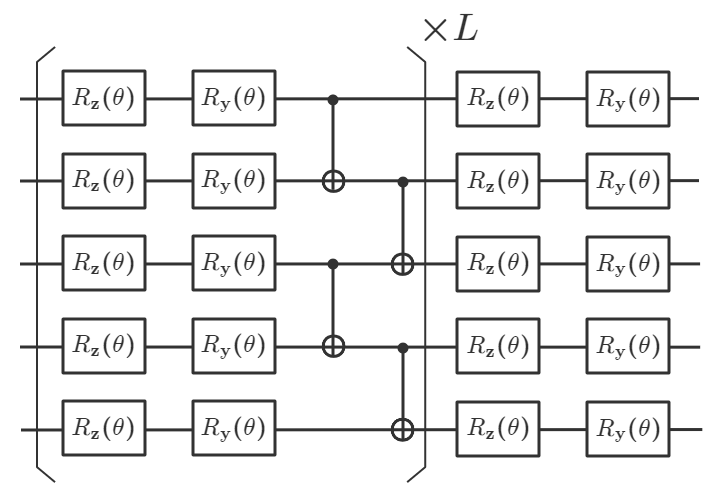}
        \label{hardware_efficient}
        \end{minipage}
    }
    \subfigure[]
    {
        \begin{minipage}{.3\linewidth}
        \centering
        \includegraphics[width=1.2\textwidth]{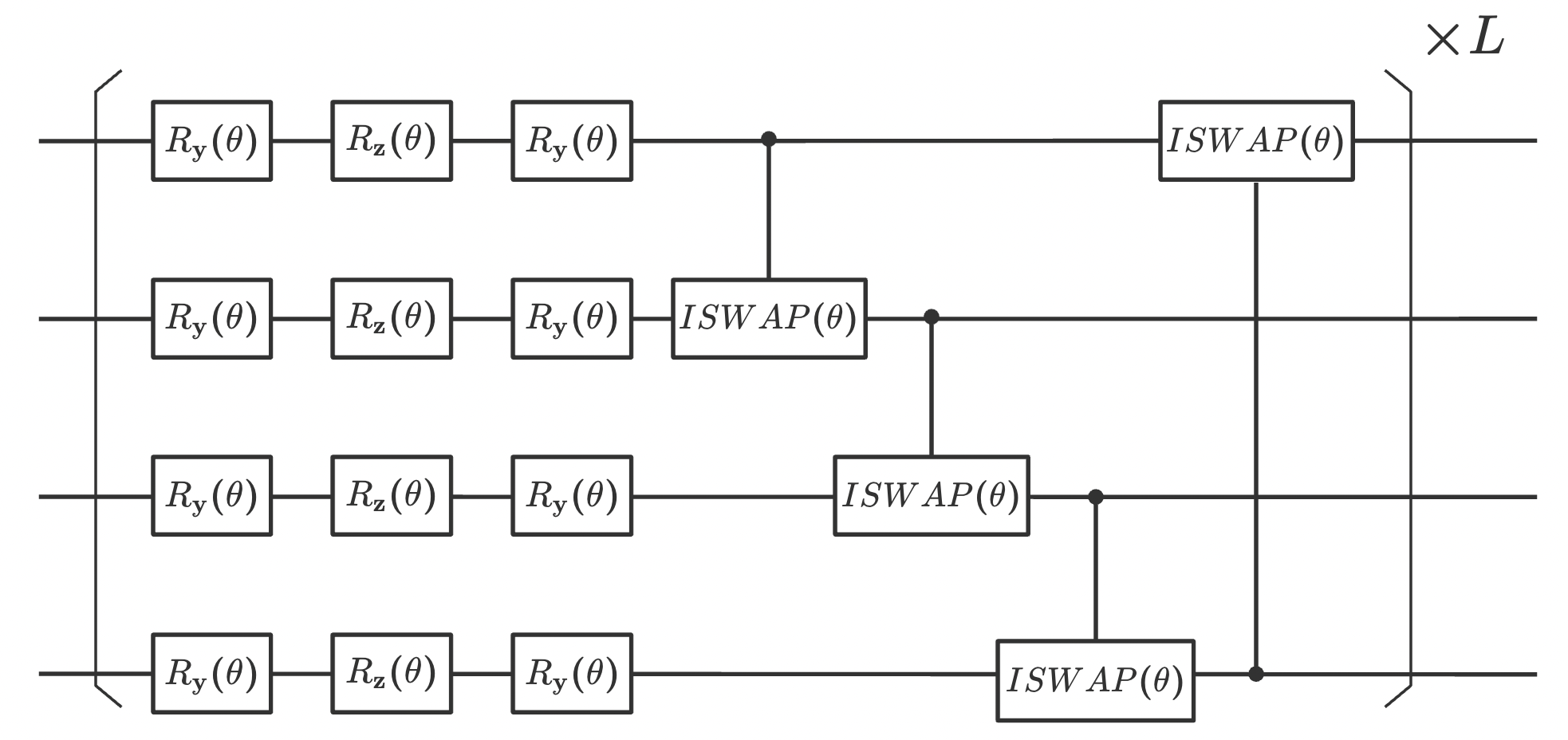}
        \label{PHE}
        \end{minipage}
    } 
    \subfigure[]
    {
        \begin{minipage}{.3\linewidth}
        \centering
        \includegraphics[width=1.2\textwidth]{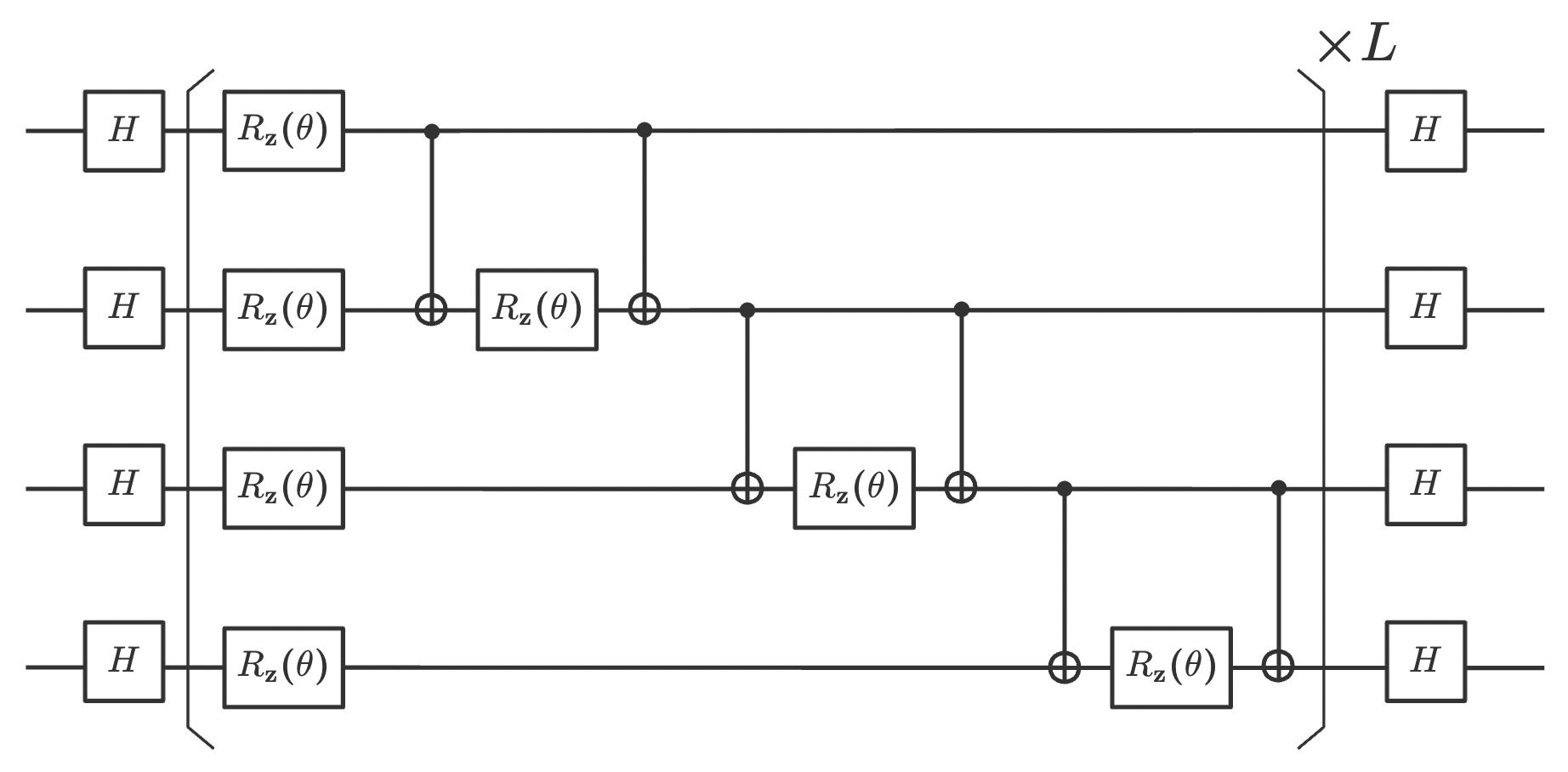}
        \label{IQP}
        \end{minipage}
    }

    \caption{Three established ansatz on which we build the VQE to solve molecules and $1$D physical models. \textbf{(a)} A $5$-qubit example of the hardware efficient ansatz with $L$ layers. \textbf{(b)} A $4$-qubit example of the parameterized two-qubit gate~(PTG) ansatz with $L$ layers. \textbf{(c)} A $4$-qubit example of the instantaneous quantum polynomial~(IQP) ansatz with $L$ layers.}
    \label{comparison_ansatz}
\end{figure}

One of the most critical applications for a variational quantum circuit ansatz is the variational quantum eigensolver~(VQE). In this section, we focus on the VQE based on the SG ansatz to solve typical molecule and $1$D quantum physical models. Furthermore, we compare the number of quantum gates required for the SG ansatz and three established ansatz in solving the VQE tasks In addition to the SG ansatz, we use the hardware-efficient~(HE) ansatz~\cite{kandala2017hardware}, the parameterized two-qubit gate~(PTG) ansatz~\cite{doi.org/10.1002/andp.202200338} and the instantaneous quantum polynomial~(IQP) ansatz~\cite{havlivcek2019supervised} to construct the VQE. As shown in figure~\ref{hardware_efficient}, HE ansatz has $L$-layer variational quantum circuits, each consisting of several single-qubit rotation and CNOT gates. PTG ansatz, shown in figure~\ref{PHE}, has $L$-layer variational quantum circuits, each of which consists of several single-qubit rotation gates and the parameterized ISWAP gates. In IQP ansatz, as shown in figure~\ref{IQP}, a Hadamard gate is applied to each qubit at the beginning and end of the circuit. In the middle of the circuit, there are $L$ layers consisting of CNOT gates and rotation Z gates. 

For the comparison, we use relative error as the indicator to characterize the performance of these ansatz. Labeling the ground energy of a given Hamiltonian $H$ as $\lambda_g$, we define a relative error as
\begin{equation}\label{relative_error}
    \epsilon(\bm{\theta})\equiv\left|\frac{\langle H(\bm{\theta})\rangle-\lambda_g}{\lambda_g}\right|\times 100\%. 
\end{equation}
where $\langle H(\bm{\theta})\rangle$ is the shorthand for the average energy obtained from measuring the quantum system as explained in equation~(\ref{E_theta}). Using the definition, we compare the number of gates required for each ansatz to achieve a threshold $\epsilon(\bm{\theta}^*)\leq 0.1\%$ after training the parameters $\bm{\theta}$. 

To find the minimum number of quantum gates for each ansatz, we use the following steps: i) choose a maximum number of layers $L_{\rm{max}}$ and set the layer number of an ansatz $L=1$; ii) train the parameters $\bm{\theta}$ through an optimization process and calculate the relative error $\epsilon(\bm{\theta}^*)$ based on the optimal parameters $\bm{\theta}^*$; iii) if $\epsilon(\bm{\theta}^*)\leq0.1\%$, we record the number of quantum gates. Otherwise, we add one more layer and repeat the training process until $L$ equals $L_{\rm{max}}$. In the meanwhile, if the ansatz with the maximum layers still fails to achieve the threshold, we will record the lowest average energy and the corresponding number of gates. We repeat this process $10$ times for each ansatz and obtain the average results. All of the simulations in the following are completed by using the python package mindquantum~\cite{mq_2021}.

\subsection{VQE for Molecule models}
We consider the VQE task to approximate the ground energy of typical molecule models. The molecules which we used for the comparison are the HF molecule, the H2O molecule and the NaH molecule. Since the Hamiltonian of these molecules is based on the fermionic model, we utilize the Jordan-Wigner transformation to transform them into the qubit model. After obtaining the qubit Hamiltonian, we calculate their FCI energies through the OpenFermion package~\cite{openfermion} and further take the FCI energies as the ground energies $\lambda_g$ for the molecule Hamiltonian. After implementing the process to find the minimum number of quantum gates as mentioned above, we summarize the average results for each ansatz in table \ref{molecule_result}. 

It can be found that the VQEs based on all of four ansatz can achieve the threshold for three molecule Hamiltonian. For convenience, we use the SG-VQE, HE-VQE, IQP-VQE and PTG-VQE to indicate the VQE based on SG ansatz, HE ansatz, IQP ansatz and PTG ansatz, respectively. For the HF molecule, SG-VQE requires an average of $57$ quantum gates to achieve the threshold. In the meanwhile, the HE-VQE requires $108$ gates, the IQP-VQE requires $101$ gates and the PTG-VQE requires $72$ gates. The SG-VQE requires $47.22\%$, $43.56\%$ $20.83\%$ fewer gates than the HE-VQE, the IQP-VQE, and the PTG-VQE. For the H2O molecule, SG-VQE requires an average of $60$ quantum gates, the HE-VQE requires $90$ gates, the IQP-VQE requires $101$ gates and the PTG-VQE requires $70$ gates. The SG-VQE requires $33.33\%$, $40.60\%$ $14.29\%$ fewer gates to achieve the threshold than the HE-VQE, the IQP-VQE, and the PTG-VQE. In the meanwhile, for the NaH molecule, SG-VQE requires an average of $93$ quantum gates, the HE-VQE requires $129$ gates, the IQP-VQE requires $120$ gates and the PTG-VQE requires $100$ gates. The SG-VQE requires $27.91\%$, $22.50\%$ $7\%$ fewer gates to achieve the threshold than the HE-VQE, the IQP-VQE, and the PTG-VQE. To sum up all of the results, we can find that the SG ansatz can significantly reduce the number of quantum gates for the VQE to solve the molecule models compared with the other three ansatz. 

\begin{table}
\caption{Simulation results to find the minimum number of quantum gates required for each ansatz to achieve the threshold $\epsilon(\bm{\theta}^*)\leq 0.1\%$ in solving the molecule models. All of the results are obtained by averaging $10$ independent simulations. In each simulation, we set the maximum number of layers to be $10$ for the ansatz. In the optimization process, we use the BFGS method to train the parameters $\bm{\theta}$ and calculate the relative error after $100$ iterations.}\label{molecule_result}

\begin{indented}
\lineup
\item[]\begin{tabular}{@{}*{5}{c}}
\br
Molecule & Ansatz  & \# Gates & $\epsilon(\bm{\theta}^*)$~$\times$~$10^{-2}~\%$  & $\epsilon(\bm{\theta}^*)\leq 0.1\%$\\
\mr
\multirow{2}{*}{HF} & SG & $57\pm 22$ & $2.590\pm0.037$ & True\\
\multirow{2}{*}{(12 qubits)} & HE & $108\pm 28$ & $2.554\pm0.048$ & True\\
& IQP & $101 \pm 32$ & $2.619\pm0.001$ & True\\
& PTG & $72 \pm 24$ & $2.620\pm0.002$ & True\\
\mr
\multirow{2}{*}{H20} & SG & $60\pm20$ & $6.660\pm0.040$ & True \\
\multirow{2}{*}{(14 qubits)}& HE & $90\pm21$ & $6.669\pm0.032$ & True\\
& IQP & $101\pm 36$ & $6.683\pm 0.001$ & True\\
& PTG & $70\pm 0$ & $6.733\pm0.109$ & True \\
\mr
\multirow{2}{*}{NaH} & SG & $93\pm46$ & $4.461\pm 1.555$ & True \\
\multirow{2}{*}{(20 qubits)} & HE & $129\pm 54$ & $5.628\pm2.154$ & True\\
& IQP &  $120\pm 0$ & $6.386\pm 2.411$ & True\\
& PTG & $100\pm 0$ & $3.976\pm 0.001$ & True \\
\br
\end{tabular}
\end{indented}
\end{table}

\subsection{VQE for 1D quantum models}
We further consider the tasks for solving the typical $1$D quantum models. The first model which we considered is the $1$D open-boundary Ising model. In general, the Hamiltonian of a $n$-qubit $1$D open-boundary Ising model is given as
\begin{equation}
    H=-J\left(\sum_{i=1}^{n-1}Z^{(i)}Z^{(i+1)}+ \gamma \sum_{i=1}^n X^{(i)}\right)
\end{equation}
where $J$ represents the inverse temperature and sets the energy scale and $\gamma$ indicates a dimensionless nearest-neighbor coupling parameter. We choose $J=1$ and $\gamma=0.5$ for our simulations. We compare the minimum number of gates required for $4$ ansatz to achieve threshold $\epsilon(\bm{\theta}^*)\leq 0.1\%$ in calculating the $15$-qubit, $20$-qubit and $24$-qubit Ising model, respectively. For each model, we use the exact diagonalization method to calculate the ground energy, and further calculate the relative error after $500$ iterations. The comparison results are summarized in table~\ref{Table:1D_Ising_result}. From the table, we can find that, for all of the three Ising models, the SG-VQE requires the fewest quantum gates to achieve the threshold. In the meanwhile, although the HE-VQE and the PTG-VQE can achieve the threshold, they require at least $25.25\%$ more quantum gates than the SG-VQE. However, the IQP-VQE fails to reach the threshold for all of the three models. 

The second model for the comparison is the Heisenberg chain. In the simulation, we focus on the task to solve the spin-$\frac{1}{2}$ XXZ model with open boundary whose Hamiltonian is given as 
\begin{equation}
    H=-J\sum_{i=1}^{n-1} \left(X^{(i)}X^{(i+1)}+Y^{(i)}Y^{(i+1)}+\gamma Z^{(i)}Z^{(i+1)}\right)
\end{equation}
where we choose $J=1$ and $\gamma=0.5$ for our simulations. We implement the comparison based on $14$-qubit, $16$-qubit and $20$-qubit Heisenberg models. For each model, we use the exact diagonalization method to calculate the ground energy, and calculate the relative error after $1500$ iterations. We summarize the simulation results in table~\ref{Table:1D_Heisenberg_result}. From the table, we can find that, for the $14$-qubit and $16$-qubit Heisenberg models, only the SG-VQE and the PTG-VQE can achieve the threshold, however, the HE-VQE and IQP-VQE fail. For these two models, the SG-VQE requires significantly fewer quantum gates than the PTG-VQE to achieve the threshold. For the $20$-qubit Heisenberg model, we can find that only the SG-VQE can achieve the threshold, and the other three fail. From all of the results based on molecule models and the $1$D quantum models, we demonstrate an advantage for the SG ansatz in significantly reducing circuit complexity and improving the effectiveness of VQE.   

\begin{table}
\caption{Simulation results to find the minimum number of quantum gates required for each ansatz to achieve the threshold $\epsilon(\bm{\theta}^*)\leq 0.1\%$ in solving the 1D Ising chains. All of the results are obtained by averaging $10$ independent simulations. In each simulation, we set the maximum number of layers to be $10$ for the ansatz. In the optimization process, we use the BFGS method to train the parameters $\bm{\theta}$ and calculate the relative error after $500$ iterations.}\label{Table:1D_Ising_result}

\begin{indented}
\lineup
\item[]\begin{tabular}{@{}*{5}{c}}
\br
Scale & Ansatz  & \# Gates & $\epsilon(\bm{\theta}^*)$~$\times$~$10^{-2}~\%$  & $\epsilon(\bm{\theta}^*)\leq 0.1\%$\\
\mr
\multirow{4}{*}{15 qubits} & SG & $99\pm 20$ & $1.482\pm1.472$ & True\\
& HE & $124\pm 28$ & $6.399\pm2.691$ & True\\
& IQP & $234 \pm 137$ & $717.7\pm0.020$ & False\\
& PTG & $210 \pm 65$ & $2.341\pm1.888$ & True\\
\mr
\multirow{4}{*}{20 qubits} & SG & $145\pm29$ & $0.552\pm0.514$ & True \\
& HE & $182\pm39$ & $3.672\pm2.501$ & True\\
& IQP & $320\pm 209$ & $686.4\pm 0.001$ & False\\
& PTG & $250\pm 67$ & $2.257\pm1.098$ & True \\
\mr
\multirow{4}{*}{24 qubits} & SG & $217\pm49$ & $0.529\pm 0.479$ & True \\
& HE & $282\pm 162$ & $2.662\pm2.041$ & True\\
& IQP &  $345\pm 151$ & $995.5\pm 397.4$ & False\\
& PTG & $324\pm 94$ & $1.642\pm 1.371$ & True \\
\br
\end{tabular}
\end{indented}
\end{table}

\begin{table}
\caption{Simulation results to find the minimum number of quantum gates required for each ansatz to achieve the threshold $\epsilon(\bm{\theta}^*)\leq 0.1\%$ in solving the 1D Heisenberg chains. All of the results are obtained by averaging $10$ independent simulations. In each simulation, we set the maximum number of layers to be $15$ for the ansatz. In the optimization process, we use the BFGS method to train the parameters $\bm{\theta}$ and calculate the relative error after $1500$ iterations.}\label{Table:1D_Heisenberg_result}
\begin{indented}
\lineup
\item[]\begin{tabular}{@{}*{5}{c}}
\br
Scale & Ansatz  & \# Gates & $\epsilon(\bm{\theta}^*)$~$\times$~$10^{-2}~\%$  & $\epsilon(\bm{\theta}^*)\leq 0.1\%$\\
\mr
\multirow{4}{*}{14 qubits} & SG & $566\pm68$ & $6.414\pm1.937$ & True\\
& HE & $475\pm125$ & $25.25\pm13.78$ & False\\
& IQP & $90\pm 17$ & $3610\pm 0.001$ & False\\
& PTG & $931\pm 99$ & $8.621\pm2.944$ & True\\
\mr
\multirow{4}{*}{16 qubits} & SG & $782\pm69$ & $6.497\pm 2.289$ & True \\
& HE & $521\pm 117$ & $46.59\pm23.94$ & False\\
& IQP & $96\pm 0$ & $3573\pm 0.001$ & False\\
& PTG & $1096\pm 80$ & $8.610\pm 1.638$ & True \\
\mr
\multirow{4}{*}{20 qubits} & SG & $1166\pm 80$ & $7.172\pm1.349$ & True \\
& HE & $728\pm 199$ & $74.81\pm47.51$ & False\\
& IQP &  $120 \pm 0$ & $3523\pm0.001$ & False\\
& PTG & $1370 \pm 155$ & $18.61\pm5.244$ & False \\
\br
\end{tabular}
\end{indented}
\end{table}

\subsection{SG-VQE for 2D quantum Ising model} 

\begin{figure}[htp]
    \centering
    
    \includegraphics[width=0.4\linewidth]{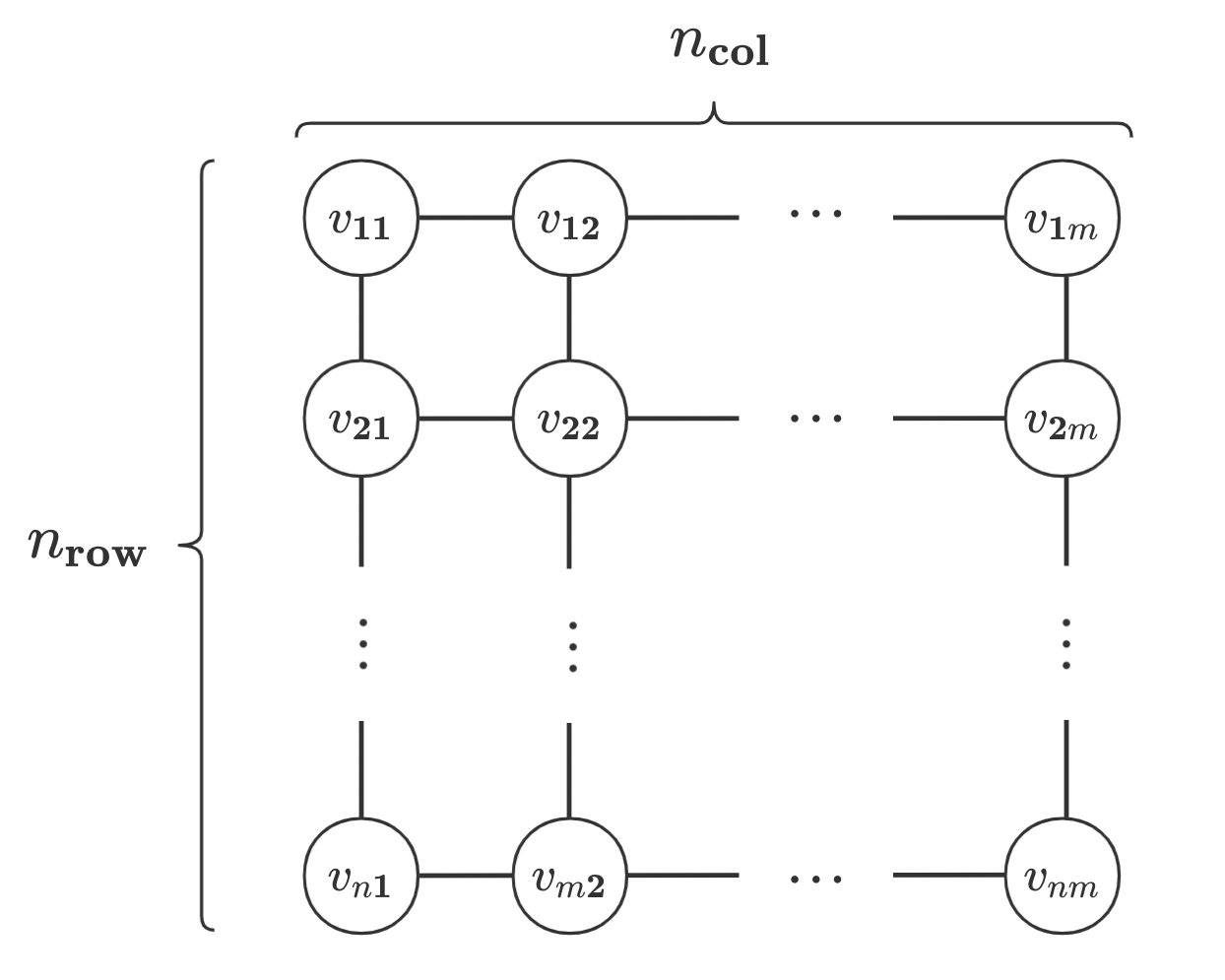}
    
    \caption{A general quantum Ising lattice with $n_{\rm{row}}$ rows and $n_{\rm{col}}$ columns. Each vertex $v_{ij}$ represents one qubit and each edge represents the nearest-neighbor interaction.}
    \label{2D_lattice}
\end{figure}

Since the SG-VQE can effectively approximate the ground energy of $1$D quantum model, in this section, we consider the case for the SG-VQE to solve the $2$D quantum Ising lattice with open boundary condition. A general quantum Ising model is shown in figure~\ref{2D_lattice} where we label the number of rows as $n_{\rm{row}}$ and the number of columns as $n_{\rm{col}}$. Hence, the total number of qubits is $n=n_{\rm{row}}n_{\rm{col}}$. The Hamiltonian of the $2$D quantum Ising model is given as
\begin{equation}\label{Ising_2D}
    H=-J\sum_{\langle i,j\rangle}Z^{(i)}Z^{(j)}-g\sum_{j}X^{(j)}
\end{equation}
where $\langle i,j\rangle$ indicates the nearest-neighbor two qubits, $g$ represents a dimensionless nearest-neighbor coupling parameter and $J$ indicates the inverse temperature and sets the energy scale. Specially, we choose $J=1$ and $g=0.5$ in our simulations.

\begin{figure}[tp]
    \centering
    
    \includegraphics[width=0.6\linewidth]{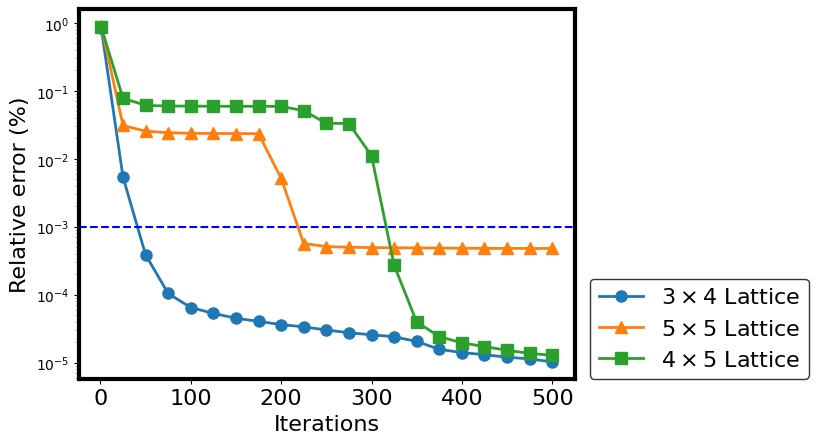}
    
    \caption{Simulation results for the SG-VQE to approximate the ground energy of $3\times 4$, $4\times 5$, $5\times 5$, three kinds of quantum Ising lattices.
    }
    \label{fig:Ising_lattice_simulation}
\end{figure}

We use the SG-VQE to solve $3\times 4$, $4\times 5$, $5\times 5$, three kinds of quantum Ising lattices. The SG ansatz is placed by using the method introduced in Section II where we first generate two lines and then apply a circuit on the qubits in each line. For the circuit, we preassign the bond dimension as $R=4$ and the number of layers to be one. As the benchmark, we use the exact diagonalization method to calculate the ground energy of each lattice model, and further take the relative error $\epsilon(\bm{\theta}^*)\leq 0.1\%$ as our optimization goal. Taking the relative error as the function of number of iterations, we summarize the simulation results in figure~\ref{fig:Ising_lattice_simulation} and the configurations for each model in table \ref{2D_simulation_results}. From figure~\ref{fig:Ising_lattice_simulation}, we can find that, after $500$ iterations, all of the SG-VQEs can achieve the threshold $\epsilon(\bm{\theta}^*)\leq 0.1\%$. In the meanwhile, it can be seen from table \ref{2D_simulation_results} that, for the $3 \times 3$ lattice, SG-VQE uses $106$ quantum gates to achieve relative error $\epsilon(\bm{\theta}^*)=0.001\%$, for the $4 \times 5$ lattice, SG-VQE uses $186$ gates to achieve relative error $\epsilon(\bm{\theta}^*)=0.0013\%$, for the $5 \times 5$ lattice, SG-VQE uses $236$ gates to achieve relative error $\epsilon(\bm{\theta}^*)=0.0479\%$. Taking these simulation results together, we show that the SG ansatz can effectively solve the quantum Ising lattice model with a few quantum gates.

\begin{table}
\caption{\label{2D_simulation_results}Configurations and simulation results for the SG-VQE to approximate the ground energy of three kinds of $2$D quantum Ising lattice. The relative error is obtained by averaging the final results of $10$ random initialization.} 

\begin{indented}
\lineup
\item[]\begin{tabular}{@{}*{6}{c}}
\br                              
$n_{\rm{row}}$ & $n_{\rm{col}}$ & \# Qubits & $g/J$ &\# Gates & $\epsilon(\bm{\theta}^*)$ \cr 
\mr
$3$ & $4$ & $12$  & $0.5$ & $106$ & $0.0010\%$\\

$4$ & $5$ & $20$ & $0.5$ & $186$ & $0.0013\%$\\

$5$ & $5$ & $25$ & $0.5$ & $236$ & $0.0479\%$\\

$3$ & $4$ & $12$  & $3.044$ & $244$ & $0.0031\%$\\

$4$ & $5$ & $20$ & $3.044$ & $744$ & $0.0021\%$\\

$5$ & $5$ & $25$ & $3.044$ & $850$ & $0.0021\%$\\ 
\br
\end{tabular}
\end{indented}
\end{table}

\begin{figure}[htp]
    \subfigure[]
    {
     \begin{minipage}{.45\linewidth}\label{Ising_open}
     \centering
      \includegraphics[width=1\textwidth]{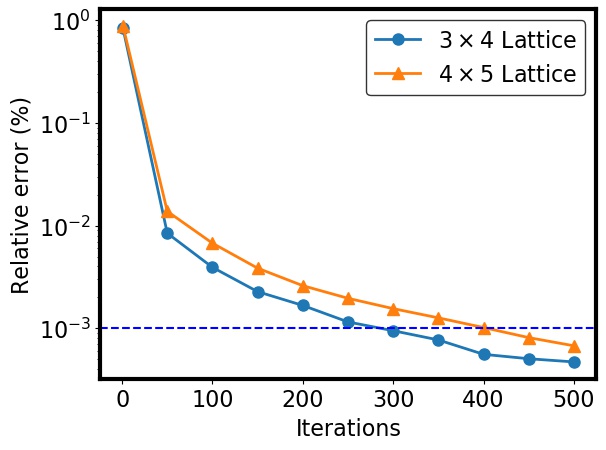}
     \end{minipage}
    }
    \subfigure[]
    {
     \begin{minipage}{.45\linewidth}\label{Ising_open_5_5}
     \centering
      \includegraphics[width=1\textwidth]{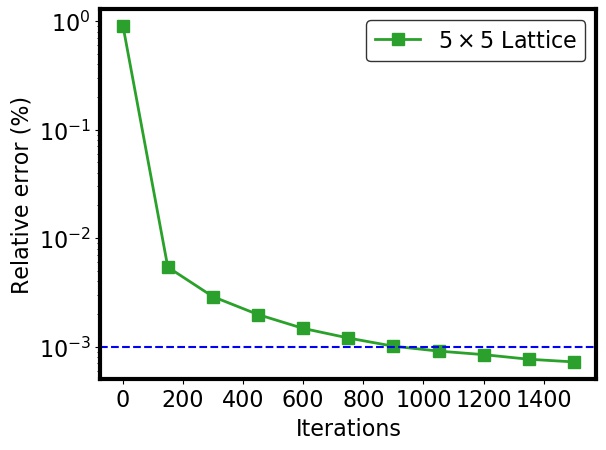}
     \end{minipage}
    } 

   \caption{ Simulation results for the SG-VQE to approximate the ground energy of three kinds of quantum Ising lattices at the quantum critical point. \textbf{(a)} Results for $3\times 4$ and $4 \times5$ lattices after $500$ iterations. \textbf{(b)} Result for $5\times 5$ lattice after $1500$ iterations.}
   \label{fig:critical_point}
\end{figure}

Besides the above simulations, we further apply the SG-VQE to solve the ground energy of the 2D quantum Ising model at the quantum critical point. The criticality of the 2D Ising model refers to the fact that the ground state stays in the boundary between the ferromagnetic phase and the antiferromagnetic phase, and typically requires much more parameters to characterize the ground state. According to Ref.~\cite{PhysRevE.66.066110}, we choose a point near the criticality, $g/J=3.044$, for $3\times 4$, $4\times 5$, $5\times 5$, three kinds of 2D quantum Ising models. Before implementing the simulations, we utilize the exact diagonalization method to calculate the ground energy for each model, and choose the relative error $\epsilon(\bm{\theta}^*)\leq 0.1\%$ as our goal. We use the SG-ansatz generated based on bond dimension $R=4$ and increase the number of layers until the relative error achieves the threshold. We summarize the simulation results in figure~\ref{fig:critical_point} and table \ref{2D_simulation_results}. From the figure, we can find that, after $500$ iterations, the SG-VQEs can achieve the threshold $\epsilon(\bm{\theta}^*)\leq 0.1\%$ for $3\times 4$ and $4\times 5$ lattices. However, it requires more iterations to achieve the threshold for the $5\times 5$ lattice. In the meanwhile, it can be obtained from table \ref{2D_simulation_results} that, for the $3 \times 3$ lattice, SG-VQE uses $244$ quantum gates to achieve relative error $\epsilon(\bm{\theta}^*)=0.0031\%$, for the $4 \times 5$ lattice, SG-VQE uses $744$ gates to achieve relative error $\epsilon(\bm{\theta}^*)=0.0021\%$, for the $5 \times 5$ lattice, SG-VQE uses $850$ gates to achieve relative error $\epsilon(\bm{\theta}^*)=0.0021\%$. Compared with the trivial case, $g/J=0.5$, the SG-VQE requires more quantum gates to achieve the threshold, but is still effective to solve the quantum Ising lattice model at the quantum critical point.

\subsection{SG-VQE for 3D quantum Ising model} 
We have shown that the SG-VQE has the ability to solve quantum Ising lattice in previous section. In this section, we consider the task for the SG-VQE to solve $3$D quantum Ising model. Similar to the $2$D Ising lattice, we use $n_{\rm{len}}$, $n_{\rm{wid}}$ and $n_{\rm{hei}}$ to indicate the number of qubits in the length, width and height of the $3$D Ising model as shown in figure~\ref{3D_model}. The number of qubits for a $3$D Ising model is equal to $n=n_{\rm{len}}n_{\rm{wid}}n_{\rm{hei}}$. The Hamiltonian of 3$D$ Ising model has the same formula as equation~\ref{Ising_2D} where we choose $J=1$ and $g=0.5$.

\begin{figure}[htp]
    \setlength{\belowcaptionskip}{-2mm}
    \centering
    
    \includegraphics[width=0.6\linewidth]{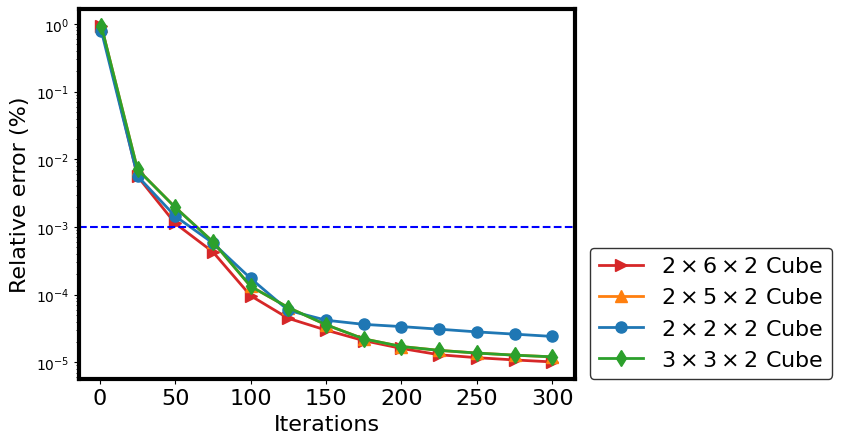}
    \caption{Simulation results for the SG-VQE to approximate the ground energy of $2\times 2\times 2$, $3 \times 3\times 2$, $2 \times 5\times 2$ and $2 \times 6\times 2$, four kinds of $3$D quantum Ising models.
    }
    \label{fig:Ising_cube_simulation}
\end{figure}

We utilize the SG-VQE to solve $2\times 2\times 2$, $3 \times 3\times 2$, $2 \times 5\times 2$ and $2 \times 6\times 2$, four kinds of $3$D quantum Ising models. The SG ansatz is designed based on the method introduced in Section II where we firstly generate several lines which traverse all of the vertexes and edges in a $3$D models and then place a circuit on the qubits in each line. We set the bond dimension $R=4$ and the number of layers to be one for all of the circuits. In the meanwhile, we calculate the ground energy of each model by using the exact diagonalization method, and take relative error $\epsilon(\bm{\theta}^*)\leq 0.1\%$ as the optimization goal. We record the change of relative error with the iteration increasing in figure~\ref{fig:Ising_cube_simulation}. The figure shows that, after $300$ iterations, all of the SG-VQEs can achieve low relative errors for all models. In the meanwhile, we record the configurations, the number of lines which we use to construct the ansatz and the final relative errors for each model in table \ref{3D_simulation_results}. It can be found from table \ref{3D_simulation_results} that, to solve the $2 \times 2 \times 2$ model, the SG-VQE uses $212$ quantum gates to achieve relative error $\epsilon(\bm{\theta}^*)=0.0024\%$, to solve the $3 \times 3 \times 2$ model, the SG-VQE uses $332$ quantum gates to achieve relative error $\epsilon(\bm{\theta}^*)=0.0012\%$, to solve the $2 \times 5 \times 2$ model, the SG-VQE uses $332$ quantum gates to achieve relative error $\epsilon(\bm{\theta}^*)=0.0012\%$ and to solve the $2 \times 6 \times 2$ model, the SG-VQE uses $339$ quantum gates to achieve relative error $\epsilon(\bm{\theta}^*)=0.0010\%$. Summarizing all of the results, we demonstrate that the SG ansatz can effectively solve the $3$D quantum Ising model with a little number of quantum gates.

\begin{table}
\caption{\label{3D_simulation_results}Configurations and simulation results for the SG-VQE to approximate the ground energy of four $3$D quantum Ising models. The relative error is obtained by averaging the final results of $10$ random initializations.} 
\begin{indented}
\lineup
\item[]\begin{tabular}{@{}*{7}{c}}
\br                              
$n_{\rm{len}}$ & $n_{\rm{wid}}$ & $n_{\rm{hei}}$ & \# Qubits & \# Lines & \# Gates & $\epsilon(\bm{\theta}^*)$ ($\times 10^{-3}$) \cr 
\mr
$2$ & $2$ & $2$ & $8$ & $4$ & $212$ & $2.4089\%$\\

$3$ & $3$ & $2$ & $18$ & $4$ & $332$ & $1.2022\%$\\

$2$ & $5$ & $2$ & $20$ & $3$ & $332$ & $1.2021\%$\\

$2$ & $6$ & $2$ & $24$ & $3$ & $339$ & $1.0114\%$\\
\br
\end{tabular}
\end{indented}
\end{table}

\section{Conclusion}\label{sec:summary}
In this work, we present an alternative variational quantum circuit ansatz, the sequentially generated ansatz. We further demonstrate that the SG ansatz can generate a MPS with polynomial circuit complexity. Our simulation results demonstrate that our circuit ansatz can be used to accurately reconstruct unknown pure and mixed quantum states which can be represented by MPSs. Furthermore, the VQE with our SG ansatz significantly reduces the circuit complexity and is more effect in solving typical molecule models and $1$D quantum models compared with several established ansatz proposals. We further numerically
demonstrate the effectiveness of SG ansatz in solving $2$D and $3$D quantum Ising models. We hope that the SG ansatz can be used for more applications in both variational quantum algorithms and quantum simulation.

\ack{Acknowledgments}
X.H. and X.W. are supported by the National Natural Science Foundation of China (Grant No.~92265208) and the National Key R\&D Program of China (Grant No.~2018YFA0306703). The authors also thank Junning Li, Shijie Pan, Qingxing Xie, Shan Jin, Shaojun Wu, Yuhan Huang for helpful discussions.

\section*{References}
\bibliographystyle{unsrt}
\bibliography{ref1}

\begin{thebibliography}{10}

\bibitem{Preskill2018quantum}
Preskill J.
\newblock Quantum {C}omputing in the {NISQ} era and beyond.
\newblock {\em Quantum}, \textbf 2:79, 2018.

\bibitem{arute2019quantum}
Arute F, Arya K, Babbush R, Bacon D, Bardin J, Barends R, Biswas R, Boixo S,
  Brandao F, Buell D, et~al.
\newblock Quantum supremacy using a programmable superconducting processor.
\newblock {\em Nature}, \textbf{574}(7779):505--510, 2019.

\bibitem{PhysRevLett.127.180501}
Wu~Y, Bao W, Cao S, Chen F, Chen M, Chen X, Chung T, Deng H, Du~Y, Fan D,
  et~al.
\newblock Strong quantum computational advantage using a superconducting
  quantum processor.
\newblock {\em Phys. Rev. Lett.}, \textbf{127}:180501, Oct 2021.

\bibitem{ZhuPan2021}
Zhu Q, Cao S, Chen F, Chen M, Chen X, Chung T, Deng H, Du~Y, Fan D, Gong M,
  et~al.
\newblock Quantum computational advantage via 60-qubit 24-cycle random circuit
  sampling.
\newblock {\em Sci. Bull.}, \textbf{67}(3):240--245, 2022.

\bibitem{zhong2020quantum}
Zhong H, Wang H, Deng Y, Chen M, Peng L, Luo Y, Qin J, Wu~D, Ding X, Hu~Y,
  et~al.
\newblock Quantum computational advantage using photons.
\newblock {\em Science}, \textbf{370}(6523):1460--1463, 2020.

\bibitem{madsen2022quantum}
Madsen L, Laudenbach F, Askarani M, Rortais F, Vincent T, Bulmer J, Miatto F,
  Neuhaus L, Helt L, Collins M, et~al.
\newblock Quantum computational advantage with a programmable photonic
  processor.
\newblock {\em Nature}, \textbf{606}(7912):75--81, 2022.

\bibitem{365700}
Shor P.
\newblock Algorithms for quantum computation: discrete logarithms and
  factoring.
\newblock In {\em Proceedings 35th Annual Symposium on Foundations of Computer
  Science}, pages 124--134, 1994.

\bibitem{childs2003exponential}
Childs A, Cleve R, Deotto E, Farhi E, Gutmann S, and Spielman D.
\newblock Exponential algorithmic speedup by a quantum walk.
\newblock In {\em Proceedings of the thirty-fifth annual ACM symposium on
  Theory of computing}, pages 59--68, 2003.

\bibitem{PhysRevLett.103.150502}
Harrow A, Hassidim A, and Lloyd S.
\newblock Quantum algorithm for linear systems of equations.
\newblock {\em Phys. Rev. Lett.}, \textbf{103}:150502, Oct 2009.

\bibitem{doi:10.1137/16M1087072}
Andrew C, Robin K, and Rolando S.
\newblock Quantum algorithm for systems of linear equations with exponentially
  improved dependence on precision.
\newblock {\em SIAM J. Comput.}, \textbf{46}(6):1920--1950, 2017.

\bibitem{PhysRevLett.109.050505}
Wiebe N, Braun D, and Lloyd S.
\newblock Quantum algorithm for data fitting.
\newblock {\em Phys. Rev. Lett.}, \textbf{109}:050505, Aug 2012.

\bibitem{PhysRevLett.113.130503}
Rebentrost P, Mohseni M, and Lloyd S.
\newblock Quantum support vector machine for big data classification.
\newblock {\em Phys. Rev. Lett.}, \textbf{113}:130503, Sep 2014.

\bibitem{lloyd2014quantum}
Lloyd S, Mohseni M, and Rebentrost P.
\newblock Quantum principal component analysis.
\newblock {\em Nat. Phys.}, \textbf{10}(9):631--633, 2014.

\bibitem{doi:10.1126/science.abn7293}
Huang H, Broughton M, Cotler J, Chen S, Li~J, Mohseni M, Neven H, Babbush R,
  Kueng R, Preskill J, and McClean R.
\newblock Quantum advantage in learning from experiments.
\newblock {\em Science}, \textbf{376}(6598):1182--1186, 2022.

\bibitem{abbas2021power}
Amira A, David S, Christa Z, Aur{\'e}lien L, Alessio F, and Stefan W.
\newblock The power of quantum neural networks.
\newblock {\em Nat. Comput. Sci.}, \textbf{1}(6):403--409, 2021.

\bibitem{PhysRevX.6.031007}
O'Malley P, Babbush R, Kivlichan D, Romero J, McClean R, Barends R, Kelly J,
  Roushan P, Tranter A, Ding N, et~al.
\newblock Scalable quantum simulation of molecular energies.
\newblock {\em Phys. Rev. X}, \textbf 6:031007, Jul 2016.

\bibitem{kandala2017hardware}
Kandala A, Mezzacapo A, Temme K, Takita M, Brink M, Chow J, and Gambetta J.
\newblock Hardware-efficient variational quantum eigensolver for small
  molecules and quantum magnets.
\newblock {\em Nature}, \textbf{549}(7671):242--246, 2017.

\bibitem{mccaskey2019quantum}
McCaskey A, Parks Z, Jakowski J, Moore S, Morris T, Humble T, and Pooser R.
\newblock Quantum chemistry as a benchmark for near-term quantum computers.
\newblock {\em npj Quantum Inf.}, \textbf 5(1):1--8, 2019.

\bibitem{mcardle2020quantum}
McArdle S, Endo S, Aspuru-Guzik A, Benjamin S, and Yuan X.
\newblock Quantum computational chemistry.
\newblock {\em Rev. Mod. Phys.}, \textbf{92}(1):015003, 2020.

\bibitem{HuangLong2022}
Huang H, Xu~X, Guo C, Tian G, Wei S, Sun X, Bao W, and Long G.
\newblock Near-term quantum computing techniques: Variational quantum
  algorithms, error mitigation, circuit compilation, benchmarking and classical
  simulation.
\newblock {\em arXiv:2211.08737}, 2022.

\bibitem{PhysRevX.7.021050}
Li~Y and Benjamin S.
\newblock Efficient variational quantum simulator incorporating active error
  minimization.
\newblock {\em Phys. Rev. X}, \textbf 7:021050, Jun 2017.

\bibitem{PhysRevLett.125.010501}
Endo S, Sun J, Li~Y, Benjamin S, and Yuan X.
\newblock Variational quantum simulation of general processes.
\newblock {\em Phys. Rev. Lett.}, \textbf{125}:010501, Jun 2020.

\bibitem{mcardle2019variational}
McArdle S, Jones T, Endo S, Li~Y, Benjamin S, and Yuan X.
\newblock Variational ansatz-based quantum simulation of imaginary time
  evolution.
\newblock {\em npj Quantum Inf.}, \textbf 5(1):1--6, 2019.

\bibitem{PRXQuantum.2.030307}
Yao Y, Gomes N, Zhang F, Wang C, Ho~K, Iadecola T, and Orth P.
\newblock Adaptive variational quantum dynamics simulations.
\newblock {\em PRX Quantum}, \textbf 2:030307, Jul 2021.

\bibitem{zhang2020low}
Zhang Z, Sun J, Yuan X, and Yung M.
\newblock Low-depth hamiltonian simulation by adaptive product formula.
\newblock {\em arXiv:2011.05283}, 2020.

\bibitem{johnson2017qvector}
Johnson P, Romero J, Olson J, Cao Y, and Aspuru-Guzik A.
\newblock Qvector: an algorithm for device-tailored quantum error correction.
\newblock {\em arXiv:1711.02249}, 2017.

\bibitem{PhysRevApplied.15.034068}
Xu~X, Benjamin S, and Yuan X.
\newblock Variational circuit compiler for quantum error correction.
\newblock {\em Phys. Rev. Applied}, \textbf{15}:034068, Mar 2021.

\bibitem{cao2022quantum}
Cao C, Zhang C, Wu~Z, Grassl M, and Zeng B.
\newblock Quantum variational learning for quantum error-correcting codes.
\newblock {\em arXiv:2204.03560}, 2022.

\bibitem{PhysRevA.98.012324}
Dallaire-Demers P and Killoran N.
\newblock Quantum generative adversarial networks.
\newblock {\em Phys. Rev. A}, \textbf{98}:012324, Jul 2018.

\bibitem{benedetti2019generative}
Benedetti M, Garcia-Pintos D, Perdomo O, Leyton-Ortega V, Nam Y, and
  Perdomo-Ortiz A.
\newblock A generative modeling approach for benchmarking and training shallow
  quantum circuits.
\newblock {\em npj Quantum Inf.}, \textbf 5(1):1--9, 2019.

\bibitem{PhysRevResearch.2.033125}
Du~Y, Hsieh M, Liu T, and Tao D.
\newblock Expressive power of parametrized quantum circuits.
\newblock {\em Phys. Rev. Res.}, \textbf 2:033125, Jul 2020.

\bibitem{PhysRevA.101.032308}
Schuld M, Bocharov A, Svore K, and Wiebe N.
\newblock Circuit-centric quantum classifiers.
\newblock {\em Phys. Rev. A}, \textbf{101}:032308, Mar 2020.

\bibitem{PhysRevA.98.032309}
Mitarai K, Negoro M, Kitagawa M, and Fujii K.
\newblock Quantum circuit learning.
\newblock {\em Phys. Rev. A}, \textbf{98}:032309, Sep 2018.

\bibitem{hou2021universal}
Hou X, Zhou G, Li~Q, Jin S, and Wang X.
\newblock A universal duplication-free quantum neural network.
\newblock {\em arXiv:2106.13211}, 2021.

\bibitem{cong2019quantum}
Cong I, Choi S, and Lukin M.
\newblock Quantum convolutional neural networks.
\newblock {\em Nat. Phys.}, \textbf{15}(12):1273--1278, 2019.

\bibitem{peruzzo2014variational}
Peruzzo A, McClean J, Shadbolt P, Yung M, Zhou X, Love P, Aspuru-Guzik A, and
  O'brien J.
\newblock A variational eigenvalue solver on a photonic quantum processor.
\newblock {\em Nat. Commun.}, \textbf{5}(1):1--7, 2014.

\bibitem{PhysRevLett.122.230401}
Parrish R, Hohenstein E, McMahon P, and Mart\'{\i}nez T.
\newblock Quantum computation of electronic transitions using a variational
  quantum eigensolver.
\newblock {\em Phys. Rev. Lett.}, \textbf {122}:230401, Jun 2019.

\bibitem{aspuru2005simulated}
Aspuru-Guzik A, Dutoi A, Love P, and Head-Gordon M.
\newblock Simulated quantum computation of molecular energies.
\newblock {\em Science}, \textbf{309}(5741):1704--1707, 2005.

\bibitem{meth2022probing}
Meth M, Kuzmin V, van Bijnen~R, Postler L, Stricker R, Blatt R, Ringbauer M,
  Monz T, Silvi P, and Schindler P.
\newblock Probing phases of quantum matter with an ion-trap tensor-network
  quantum eigensolver.
\newblock {\em arXiv:2203.13271}, 2022.

\bibitem{PhysRevX.8.031022}
Hempel C, Maier C, Romero J, McClean J, Monz T, Shen H, Jurcevic P, Lanyon B,
  Love P, Babbush R, Aspuru-Guzik A, Blatt R, and Roos C.
\newblock Quantum chemistry calculations on a trapped-ion quantum simulator.
\newblock {\em Phys. Rev. X}, \textbf 8:031022, Jul 2018.

\bibitem{PhysRevA.95.020501}
Shen Y, Zhang X, Zhang S, Zhang J, Yung M, and Kim K.
\newblock Quantum implementation of the unitary coupled cluster for simulating
  molecular electronic structure.
\newblock {\em Phys. Rev. A}, \textbf{95}:020501(R), Feb 2017.

\bibitem{Lee:22}
Lee D, Lee J, Hong S, Lim H, Cho Y, Han S, Shin H, Rehman J, and Kim Y.
\newblock Error-mitigated photonic variational quantum eigensolver using a
  single-photon ququart.
\newblock {\em Optica}, \textbf 9(1):88--95, Jan 2022.

\bibitem{li2011solving}
Li~Z, Yung M, Chen H, Lu~D, Whitfield J, Peng X, Aspuru-Guzik A, and Du~J.
\newblock Solving quantum ground-state problems with nuclear magnetic
  resonance.
\newblock {\em Sci. Rep.}, \textbf 1(1):1--8, 2011.

\bibitem{PhysRevX.8.011021}
Colless J, Ramasesh V, Dahlen D, Blok M, Kimchi-Schwartz M, McClean J, Carter
  J, de~Jong~W, and Siddiqi I.
\newblock Computation of molecular spectra on a quantum processor with an
  error-resilient algorithm.
\newblock {\em Phys. Rev. X}, \textbf{8}:011021, Feb 2018.

\bibitem{Sweke2020stochasticgradient}
Sweke R, Wilde F, Meyer J, Schuld M, Faehrmann P, Meynard-Piganeau B, and
  Eisert J.
\newblock Stochastic gradient descent for hybrid quantum-classical
  optimization.
\newblock {\em {Quantum}}, \textbf{4}:314, August 2020.

\bibitem{suzuki2021normalized}
Suzuki Y, Yano H, Raymond R, and Yamamoto N.
\newblock Normalized gradient descent for variational quantum algorithms.
\newblock In {\em 2021 IEEE International Conference on Quantum Computing and
  Engineering (QCE)}, pages 1--9. IEEE, 2021.

\bibitem{mcclean2018barren}
McClean J, Boixo S, Smelyanskiy V, Babbush R, and Neven H.
\newblock Barren plateaus in quantum neural network training landscapes.
\newblock {\em Nat. Commun.}, \textbf{9}(1):1--6, 2018.

\bibitem{wang2021noise}
Wang S, Fontana E, Cerezo M, Sharma K, Sone A, Cincio L, and Coles P.
\newblock Noise-induced barren plateaus in variational quantum algorithms.
\newblock {\em Nat. Commun.}, \textbf{12}(1):1--11, 2021.

\bibitem{Arrasmith2021effectofbarren}
Arrasmith A, Cerezo M, Czarnik P, Cincio L, and Coles P.
\newblock Effect of barren plateaus on gradient-free optimization.
\newblock {\em Quantum}, \textbf{5}:558, October 2021.

\bibitem{PRXQuantum.3.020365}
Sack S, Medina R, Michailidis A, Kueng R, and Serbyn M.
\newblock Avoiding barren plateaus using classical shadows.
\newblock {\em PRX Quantum}, \textbf 3:020365, Jun 2022.

\bibitem{grimsley2022adapt}
Grimsley H, Barron G, Barnes E, Economou S, and Mayhall N.
\newblock Adapt-vqe is insensitive to rough parameter landscapes and barren
  plateaus.
\newblock {\em arXiv:2204.07179}, 2022.

\bibitem{Grant2019initialization}
Grant E, Wossnig L, Ostaszewski M, and Benedetti M.
\newblock An initialization strategy for addressing barren plateaus in
  parametrized quantum circuits.
\newblock {\em {Quantum}}, \textbf{3}:214, December 2019.

\bibitem{doi:10.1021/acs.jctc.8b01004}
Lee J, Huggins W, Head-Gordon M, and Whaley K.
\newblock Generalized unitary coupled cluster wave functions for quantum
  computation.
\newblock {\em J. Chem. Theory Comput.}, \textbf{15}(1):311--324, 2019.

\bibitem{a12020034}
Hadfield S, Wang Z, O'Gorman B, Rieffel E, Venturelli D, and Biswas R.
\newblock From the quantum approximate optimization algorithm to a quantum
  alternating operator ansatz.
\newblock {\em Lect. Notes. Comput. Sc.}, \textbf{12}(2), 2019.

\bibitem{SchuchCirac2008}
Schuch N, Wolf M, Verstraete F, and Cirac J.
\newblock Simulation of quantum many-body systems with strings of operators and
  monte carlo tensor contractions.
\newblock {\em Phys. Rev. Lett.}, \textbf{100}:040501, Jan 2008.

\bibitem{GlasserCirac2018}
Glasser I, Pancotti N, August M, Rodriguez I, and I~Cirac.
\newblock Neural-network quantum states, string-bond states, and chiral
  topological states.
\newblock {\em Phys. Rev. X}, \textbf 8:011006, Jan 2018.

\bibitem{9605301}
Suzuki Y, Yano H, Raymond R, and Yamamoto N.
\newblock Normalized gradient descent for variational quantum algorithms.
\newblock In {\em 2021 IEEE International Conference on Quantum Computing and
  Engineering (QCE)}, pages 1--9, Los Alamitos, CA, USA, 2021. IEEE Computer
  Society.

\bibitem{liu1989limited}
Liu D and Nocedal J.
\newblock On the limited memory bfgs method for large scale optimization.
\newblock {\em Math. Program.}, \textbf{45}(1):503--528, 1989.

\bibitem{kingma2014adam}
Kingma D and Ba~J.
\newblock Adam:a method for stochastic optimization.
\newblock {\em arXiv:1412.6980}, 2014.

\bibitem{hastings2007area}
Hastings M.
\newblock An area law for one-dimensional quantum systems.
\newblock {\em J. Stat. Mech. Theory Exp.}, \textbf{2007}(08):P08024, 2007.

\bibitem{cramer2010efficient}
Cramer M, Plenio M, Flammia S, Somma R, Gross D, Bartlett S, Landon-Cardinal O,
  Poulin D, and Liu Y.
\newblock Efficient quantum state tomography.
\newblock {\em Nat. Commun.}, \textbf{1}(1):1--7, 2010.

\bibitem{LiuWu2020}
Liu Y, Wang D, Xue S, Huang A, Fu~X, Qiang X, Xu~P, Huang H, Deng M, Gu~C, Yang
  X, and Wu~J.
\newblock Variational quantum circuits for quantum state tomography.
\newblock {\em Phys. Rev. A}, \textbf{101}:052316, May 2020.

\bibitem{doi.org/10.1002/andp.202200338}
Rasmussen S and Zinner N.
\newblock Parameterized two-qubit gates for enhanced variational quantum
  eigensolver.
\newblock {\em Ann. Phys.}, 534:2200338, 2022.

\bibitem{havlivcek2019supervised}
Havl{\'\i}{\v{c}}ek V, C{\'o}rcoles A, Temme K, Harrow A, Kandala A, Chow J,
  and Gambetta J.
\newblock Supervised learning with quantum-enhanced feature spaces.
\newblock {\em Nature}, \textbf{567}(7747):209--212, 2019.

\bibitem{mq_2021}
MindQuantum Developer.
\newblock Mindquantum, version 0.6.0, March 2021.

\bibitem{openfermion}
McClean J, Rubin N, Sung K, Kivlichan I, Bonet-Monroig X, Cao Y, Dai C, Fried
  S, Gidney C, Gimby B, et~al.
\newblock Openfermion: the electronic structure package for quantum computers.
\newblock {\em Quantum Sci. Technol.}, \textbf{5}(3):034014, 2020.

\bibitem{PhysRevE.66.066110}
Bl\"ote H and Deng Y.
\newblock Cluster monte carlo simulation of the transverse ising model.
\newblock {\em Phys. Rev. E}, \textbf{66}:066110, Dec 2002.

\end{thebibliography}

\end{document}